\documentclass[modern]{aastex63}
\usepackage[utf8]{inputenc}
\usepackage{graphicx}
\usepackage{natbib}
\usepackage{amsmath}
\usepackage{hyperref}
\usepackage{listings}

\shorttitle{Updated YSO models}
\shortauthors{Richardson et al.}

\begin{document}
\title{An updated modular set of synthetic spectral energy distributions for young stellar objects}
\author{Theo Richardson}
\affiliation{Department of Astronomy, University of Florida, P.O. Box 112055, Gainesville, FL 32611}
\email{terichard57@gmail.com}

\author{Adam Ginsburg}
\affiliation{Department of Astronomy, University of Florida, P.O. Box 112055, Gainesville, FL 32611}

\author{R\'{e}my Indebetouw}
\affiliation{National Radio Astronomy Observatory, 520 Edgemont Road Charlottesville, VA 22903; rindebet@nrao.edu}
\affiliation{Department of Astronomy, University of Virginia, P.O. Box 3818, Charlottesville, VA 22903-0818; remy@virginia.edu}

\author{Thomas P. Robitaille}
\affiliation{Aperio Software, Insight House, Riverside Business Park, Stoney Common Road, Stansted, CM24 8PL, United Kingdom; thomas.robitaille@gmail.com}

\begin{abstract}
Measured properties of young stellar objects (YSOs) are key tools for research into pre-main-sequence stellar evolution. YSO properties are commonly measured by comparing observed radiation to existing grids of template YSO spectral energy distributions (SEDs) calculated by radiative transfer. These grids are often sampled and constructed using simple models of mass assembly/accretion over time. However, because we do not yet have a complete theory of star formation, the choice of model sets the tracked parameters and range of allowed values. By construction, then, the assumed model limits the measurements that can be made using the grid. Radiative transfer models not constrained by specific accretion histories would enable assessment of a wider range of theories. We present an updated version of the \citet{robitaille2017} set of YSO SEDs, a collection of models with no assumed evolutionary theory. We outline our newly calculated properties: envelope mass, weighted-average dust temperature, disk stability, and circumstellar $A_{\rm V}$. We also convolve the SEDs with new filters, including JWST, and provide users the ability to perform additional convolutions. We find a correlation between the average temperature and millimeter-wavelength brightness of optically thin dust in our models and discuss its ramifications for mass measurements of pre- and protostellar cores. We also compare the positions of YSOs of different observational classes and evolutionary stages in IR color space and use our models to quantify the extent to which class and stage may be confused due to observational effects. Our updated models are released to the public.
\end{abstract}

\section{Introduction}\label{sec:1}
Many theories have been developed to explain the process of star formation from a collapsing core. Among these are spherically symmetric collapse \citep{shu1977} modified by including turbulence \citep{mckee2002,mckee2003}, competitive accretion \citep{bonnell1997,bonnell2001}, stellar collision \citep{bally2005}, and more recent variants that fall between these \citep[ex.][]{mckee2010,offner2011,myers2014,padoan2020}.

The most direct way to constrain these mechanics is to observe stellar precursors, young stellar objects (YSOs), and determine their properties and physical states. A common tool for making these measurements, particularly in cases where the YSOs are unresolved, is a grid of SED models that has been pre-computed using radiative transfer simulations. Properties of an observed YSO can then be measured via fitting the measured SED to these template models. 

Given the number of theories that exist to explain the process of star formation, many such model grids have been made for the purpose of measuring YSO properties \citep[ex.][]{robitaille2006,robitaille2007,furlan2016,haworth2018,zhang2018}.
However, these grids often face common limitations. By construction, grids typically assume a particular theory of star formation (i.e. a particular accretion history and expected envelope/disk/surrounding mass density distribution) that influences the parameters included in the models and the area of parameter space sampled. Models in these grids also often span small regions of this parameter space or are purposefully sampled to prioritize particular combinations of values which are deemed ``realistic" by the underlying theory. At times, as in the case of \citet[][R06]{robitaille2006}, the models in a grid may all be very similar (e.g. having the same basic components of a central source, a disk, a rotationally flattened envelope, and outflow cavities), varying only in physical parameters that have minimal effect on observables. These practices limit the extent to which a grid of SED models may be used to accurately determine the properties of an observed YSO. Moreover, building a model grid with the assumption of a particular theory removes the ability to use the grid as a tool to test the theory, as it will be implicit in all results obtained through use of the grid.

The SED models presented in \citet[][R17]{robitaille2017} remove these limitations by providing a large set of models spanning a number of different geometries (defined by the presence or absence of features like disks or outflow cavities) each of which are shaped by a common set of randomly sampled parameters. These choices make this set widely applicable as a tool for measuring YSO properties that is, by design, agnostic to accretion history and stellar evolution model. However, the physical parameters of each YSO in R17 are limited to those that are required for the calculation of an SED via radiative transfer. There are additional properties that can be derived for each model that can facilitate the interpretation of observations. As an example, R17 does not include the mass present in the circumstellar envelope of each model YSO (i.e. the core mass). In many theories, a core mass uniquely maps to a final stellar mass with a fixed efficiency \citep[ex.][]{mckee2002,mckee2003,federrath2012}. The core mass is therefore a sought-after observable quantity to test such theories and is often a crucial component in drawing larger conclusions about the mechanics of star formation \citep[ex.][]{almaimf1,almaimf2,almaimf3}. The randomly sampled properties can also give rise to YSO models that are not consistent with any theory of star formation or are otherwise unphysical; for example, some central sources exist below the stellar main sequence or have disks with inner radii larger than their outer radii. These ``unrealistic" models can be difficult to identify based only on the information included in R17. Finally, while the SEDs in R17 have been convolved with filters from several IR instruments, the advent of JWST brings a wealth of new data that the models could be used to interpret. The existing convolutions also do not make use of the full wavelength range of the R17 SEDs, when these models may also be useful as a point of comparison for non-IR observers.

In this paper, we present a substantial update to the R17 SED models that addresses these limitations. Section \ref{sec:2} contains an overview of the models circa 2017. In Section \ref{sec:3} we outline our additions to the models, and in Section \ref{sec:4} we demonstrate how our additions may be used to improve measurements of YSO properties. We make concluding remarks in Section \ref{sec:5}.

\section{The Robitaille (2017) Models}\label{sec:2}
In this section, we provide an overview of the state of the YSO SED models as released alongside R17. A comprehensive presentation of the models is available in the companion paper; our focus is instead on details either directly relevant to this work or useful in understanding the structure of the model set.

\subsection{Geometries}\label{sec:2.1}
\begin{figure}
    \centering
    \includegraphics{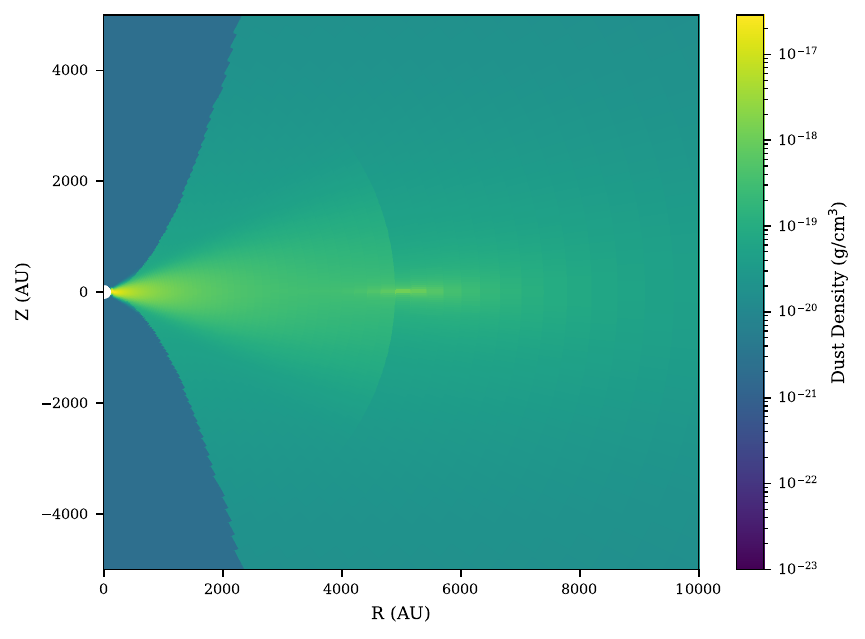}
    \caption{The dust density profile of an R17 model following \citet{ulrich1976}, an axisymmetric profile resulting from rotational flattening of an envelope and infalling material \eqref{eq:ulrich}. The model is chosen from the most populated `\texttt{spubhmi}' geometry as a representative of the most common shape of modeled YSO. This provides a sample visualization of bipolar cavities, a passive disk, and the overdensity resulting from mass inflow at the outer edge of an accretion disk (or alternatively, the centrifugal radius) as put forth in U76.}
    \label{fig:densityprofile}
\end{figure}

Models in R17 are grouped by geometry, defined by the presence or absence of dust density structures. A full list of the free parameters that shape the models, corresponding value ranges, and effect on the resulting geometry can be found in Tables 1 and 2 of R17. Broadly, the geometries are defined as a source plus zero or more features, which may include envelopes, disks, bipolar cavities, and an ambient medium. In the following subsections, we provide a short discussion of how these features are implemented, and point to the appropriate sections of R17 for those interested in greater detail. A visualization of a model with all of these features present is plotted in Figure \ref{fig:densityprofile}.

\subsubsection{Envelopes}\label{sec:2.1.1}
Dust in every R17 envelope follows one of two density profiles. The majority of models have a rotationally flattened envelope as prescribed by \citet{ulrich1976}, hereinafter U76. This profile is given by the following equation:
\begin{equation}
    \rho(r,\theta)=\rho_{0}^{\rm env}\left(\frac{r}{R_{\rm c}}\right)^{-3/2}
    \left(1+\frac{\mu}{\mu_{0}}\right)^{-1/2}\left(\frac{\mu}{\mu_{0}}+\frac{2\mu_{0}^{2}R_{\rm c}}{r}\right)^{-1}
    \label{eq:ulrich}
\end{equation}
In \eqref{eq:ulrich}, $\rho_{0}^{\rm env}$ is a density scale defined as:
\begin{equation}
    \rho_{0}^{\rm env} = \frac{\dot{M}_{\rm env}}{4\pi\left(GM_\star R_{\rm c}^3\right)^{1/2}}
    \label{eq:rho_env}
\end{equation}
where $R_{\rm c}$ is the ``centrifugal radius'' where infalling material piles up due to angular momentum, $\mu\equiv\cos\theta$, and $\mu_{0}$ is the cosine of the initial polar angle of the streamline intersecting the point $(r,\theta)$. The remainder of envelopes have simple power-law density profiles:
\begin{equation}
    \rho(r)=\rho_{0}^{\rm env}\left(\frac{r}{r_{0}}\right)^{\gamma}
    \label{eq:powerlaw}
\end{equation}
where $\rho_{0}^{\rm env}$ is the density defined at $r_{0}=1000$ AU and $\gamma$ the constant power-law exponent. In the model names, the U76 and power-law envelope density profiles are respectively denoted `u' and `p' as the third character in the geometry identifier. For more details, see Section 3.2.3 of R17.

\newpage
\subsubsection{Disks}\label{sec:2.1.2}
Envelopes with a U76 density profile may also have an accompanying flared disk (denoted `p' as the second character in the geometry identifier.) The density profile of the disk is as follows:
\begin{equation}
    \rho(R,z,\phi)=\rho_{0}^{\rm disk}\left(\frac{R_{0}}{R}\right)^{\beta-p}\exp\left[-\frac{1}{2}\left(\frac{z}{h(R)}\right)^{2}\right]
\end{equation}
where the disk scale height $h_{0}$ defines the corresponding polar radius $R_{0}$ and $h(R)$:
\begin{equation}
    h(R) = h_{0}\left(\frac{R}{R_{0}}\right)^{\beta}
\end{equation}
The disk mass that determines $\rho_{0}^{\rm disk}$, surface density profile $p$, flaring power $\beta$, spatial extent $R_{\rm min}^{\rm disk}$/$R_{\rm max}^{\rm disk}$, and scale height $h_0$ are free parameters that are assigned randomly when constructing a model.

We note that by construction, disks are intended to be passive (i.e. non-accreting and heated solely by the central source.) The impact of ignoring accretion onto a YSO's central source is minimal, or degenerate with an increase in luminosity of said source. Ignoring accretion within the disk itself does have a larger impact on the shorter-wavelength parts of the model SEDs down to the MIR and particularly in the UV regime. For more details, see Section 3.2.2 of R17.

\subsubsection{Cavities}\label{sec:2.1.3}
Many envelopes in R17 include bipolar cavities (denoted `b' as the fourth character in the geometry identifier) that replace parts of the envelope with a constant, lower density $\rho_{0}^{\rm cav}$. The cavity walls are defined by the relationship between distance from the $\mu=0$ plane $z$ and distance from the z-axis $R$:
\begin{equation}
    z(R)=R_{0}\cos\theta_{0}\left(\frac{R}{R_{0}\sin\theta_{0}}\right)^{c}
\end{equation}
where $R_{0}$ is set to 10,000 AU, $c$ is the cavity power, and $\theta_{0}$ is the angle to the cavity wall from the $z$-axis at $R_{0}$. The lower-density cavity persists out to the radius where the surrounding envelope reaches $\rho_{0}^{\rm cav}$. For more details, see Section 3.2.4 of R17.

\subsubsection{Ambient medium}\label{sec:2.1.4}
An ambient medium is included in the models to simulate the presence of an interstellar medium. It acts as a lower limit to the dust density in geometries that include it (denoted `m' as the sixth character in the geometry identifier.) Every model with an envelope also has an ambient medium, while for models that are only a star or a star and disk, the medium may or may not be present. Dust in the medium is always set to a density of $10^{-23}$ g/cm$^3$ and a temperature of 10 K. For more details, see Section 3.2.5 of R17.

\subsection{Dust}\label{sec:2.2}
All models in R17 utilize a dust model from \citet[][D03]{draine2003a,draine2003b} with the \citet{weingartner2001} Milky Way grain size distribution A for $R_{\rm V}$ = 5.5 and carbon abundance $C/H$ renormalized to 42.6 ppm. Mie scattering properties were calculated using a modified \texttt{bhmie} routine from \citet{bohren1983}. 

A large component of the work we present in this paper is concerned with the dust mass present in a YSO and our ability to measure it accurately (Sections \ref{sec:3.1}, \ref{sec:4.1}) We therefore require the opacity to absorption $\kappa_\nu$ of this dust. In the file containing the dust model released alongside R17, the opacity to extinction $\chi_\nu$ and albedo of the dust $a$ are included. $\kappa_\nu$ can be obtained by subtracting the scattering component from $\chi_\nu$, as follows:

\begin{equation}
    \kappa_\nu = \chi_\nu \times \left(1-a\right)
    \label{eq:kappa}
\end{equation}

\begin{figure}
    \centering
    \includegraphics{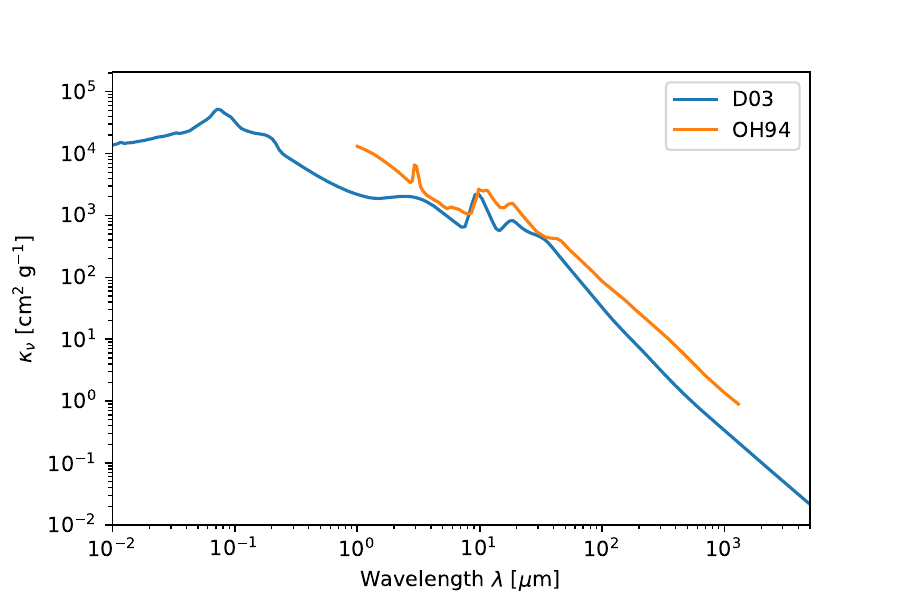}
    \caption{The opacity to absorption of dust from \citet{draine2003a,draine2003b}, $\kappa_\nu$, used in the R17 models. Opacities from \citet{ossenkopf1994} for dust grains with thin ice mantles are plotted for comparison, assuming a density of 10$^6$ cm$^{-3}$ after 10$^5$ years of coagulation.}
    \label{fig:opacities}
\end{figure}

The resulting opacities used for our comparison are plotted in Figure \ref{fig:opacities}. We also plot a set of commonly used dust opacities from \citet[][OH94]{ossenkopf1994} for comparison. The OH94 opacities are for dust grains with thin ice mantles, coagulated at 10$^6$ cm$^{-3}$ after 10$^5$ years. Dust from OH94 is generally more opaque; the values are greater by a factor of $\approx$4 than the dust from D03 at wavelengths longer than 100 $\mu$m.

\subsection{SEDs}\label{sec:2.3}
All models in R17 are associated with a suite of SEDs. Flux values for each SED are derived from the simulated radiation of the YSO as viewed within 20 log-spaced apertures between $10^{2}$-$10^{6}$ AU. All SEDs span a wavelength range of 0.01 $\mu$m to 5 mm evaluated at 200 log-spaced wavelengths. Models with density profiles dependent on the $\theta$ coordinate (i.e. models with cavities/disks) have multiple associated SEDs calculated at different inclinations. These viewing angles have been randomly sampled within $10^{\circ}$ increments from $0^{\circ}-90^{\circ}$ to preserve even coverage. All flux values are normalized to a distance of 1 kpc.

SEDs were calculated by running \texttt{Hyperion}, an open-source Monte Carlo radiative transfer code\footnote{In the data released alongside R17, some models did not have associated SEDs; a fraction of the \texttt{Hyperion} runs were not completed with the resources allotted to them. We discuss this further in Section \ref{sec:3.6}.} \citep{robitaille2011}.

\section{Additions}\label{sec:3}
In this section, we provide an outline of the additions we make to the R17 models.
\subsection{Mass calculation}\label{sec:3.1}
In R17, the spatial extent and density distribution of each protostellar envelope is specified, but the emergent parameter of mass in the envelope is not included. However, this is an important property that is often sought out and would therefore be useful information for those looking to constrain YSO properties using these models. Knowing the mass also opens up more utility for the models in other areas; for example, quantifying the relationship between the flux measured from a core and its mass, as we do in Section \ref{sec:4.1}. We are therefore interested in making the calculation of mass in the envelope of each model.

All R17 models are defined on a $400\times300$ spherical grid of cells in radius $r$ and polar angle $\theta$. All models are constructed to be axisymmetric, so no $\phi$ variation is required. Cells are log-spaced in radius over the interval ($R_{\rm env,min}$,$\sqrt{2}R_{\rm env,max}$) and roughly linearly spaced in angle over ($-\pi,\pi$). $R_{\rm env,min}$ is the inner radius of an envelope (or disk) and $R_{\rm env,max}$ is the outer extent of the model. $R_{\rm env,min}$ is set either at the dust sublimation radius around the source, $R_{\rm sub}$, or a randomly sampled value; this is determined through the model geometry. $R_{\rm sub}$ is model-dependent; it is defined as the radius where optically thin dust in the envelope would reach a temperature of 1600 K when heated by the central source in each model. $R_{\rm env,max}$ is not explicitly defined in R17, but is functionally either the radius where dust in the non-ambient density structures no longer exceeds $\rho_{\rm ISM}$ (defined as $10^{-23}$ g/cm$^{3}$ across all models) or where the optically thin dust temperature reaches the ambient temperature of 10 K, whichever is larger.

We calculate the mass present in each model by integrating the dust density profile. We perform integration using the cells as differential elements; we multiply the volume of each cell with the density defined at its center in log space to calculate the mass within each. Cell centers and sizes for each model are imported from the output of the associated \texttt{Hyperion} run. The SEDs are defined in a series of radial apertures; we calculate the mass contained within spheres of those radii, $M_{\rm core,dust}(<R)$, by identifying the cells within each radius and summing their masses. To ensure that material outside the envelope does not contribute to the mass, we only consider cells within the radius where any non-ambient dust density structure exceeds the ambient density. This treatment of numerical integration allows us to calculate masses in a way that is consistent with the setup for \texttt{Hyperion} without overweighting an ambient medium. It is also flexible enough to account for the multiple geometries present in R17, as the grid of cells is able to resolve the features of each model by construction.

Our method of integration successfully calculates the dust mass present within the density structures of each YSO. To extend that to the total mass, we assume a constant gas-to-dust ratio (GDR) of 100; this is a common approximation to the metallicity- and distance-dependent values found in \citet{draine2007} \citep[ex.][]{elia2013,konig2017}. The mass values we use and report hereinafter are therefore total mass values where a GDR has been assumed, as opposed to dust masses directly calculated from the R17 models. We have calculated $M_{\rm core, total}(<R)$ for each model within every aperture in which the SEDs are defined.

In addition to the mass contained in spherical apertures around the central source of each YSO, we have calculated an alternate set of masses that sum cells along the line of sight (i.e. within a cylindrical viewing aperture centered on the source as opposed to within a spherical radius from the source.) For these masses, we maintain the same set of radii. These results are intended to serve as ``observable" masses, as this treatment corresponds more closely with observations of YSOs, which occur along external sightlines within some viewing aperture. This approach is also more consistent with how fluxes are calculated in \texttt{Hyperion}, where the photons received within a given circular aperture along a sightline are summed. Per section 4.2 of R17, each model SED in the set has been background-subtracted within each viewing aperture, where the background consists of a slab of dust with density $10^{-23}$ g/cm$^3$, thickness of 2 $\times R_{\rm max}$ for the model, and emitting modified blackbody radiation at 10 K. To ensure we track only material contributing to the model SEDs, we do not include cells at the ambient density in this set of mass calculations either, effectively ``background-subtracting" the calculated masses.

For our masses within spherical radii, we set the mass within apertures that exceed the extent of the model to NaN as an indication of which apertures are out of bounds.

\subsection{Temperatures}\label{sec:3.2}
\begin{figure}
    \centering   
    \includegraphics[scale=0.9]{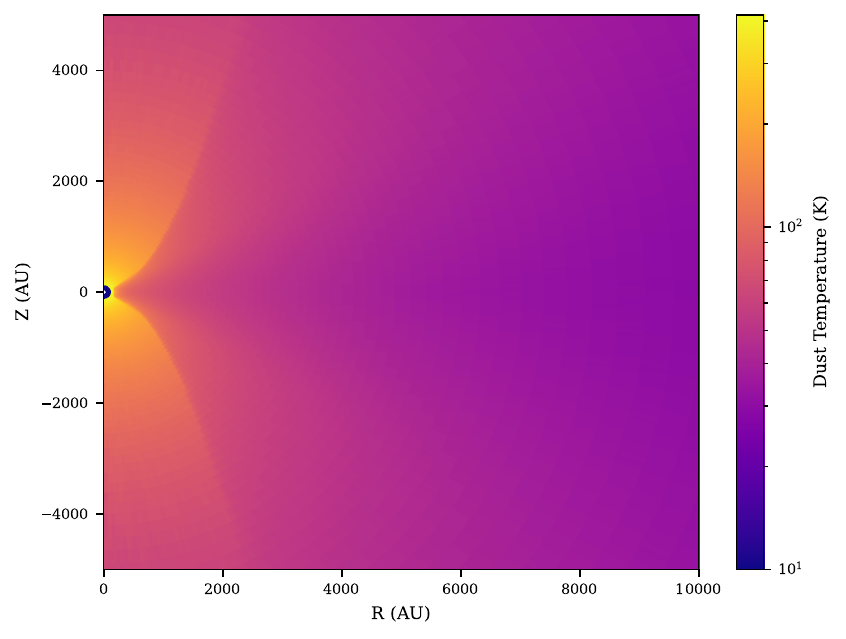}
    \caption{The same as Figure \ref{fig:densityprofile}, but colored using the dust temperature in each cell. Combining this with the mass contained in each cell allows calculation of the mass-weighted average dust temperature within a series of radial apertures. The model geometry, \texttt{spubhmi}, features rotationally flattened envelopes with bipolar cavities and passive disks.}
    \label{fig:tempprofile}
\end{figure}

\texttt{Hyperion}, as part of a run, calculates the specific energy absorbed by the dust in each cell for each model. This was used in R17 to create temperature profiles for each model. An example of such a temperature profile can be seen in Figure \ref{fig:tempprofile}. In the course of this work, we have used these profiles to calculated mass-weighted average temperatures within each viewing aperture. Each cell in the grid on which a model is defined has an associated temperature. To calculate the mass-weighted average, we sum the products of the temperatures and masses of all cells within an aperture and divide by the total mass in the aperture. As in Section \ref{sec:3.1}, temperatures are calculated both in spheres around the source and in cylinders along the line of sight. Material at the background density is not included in the calculations.

\subsection{Disk properties}\label{sec:3.3}
Many models in R17 include a passive disk (with properties given in Section \ref{sec:2.1.2}). Here as elsewhere in the models, the randomly sampled parameters may result in unrealistic disks in the sense that they may be too unstable to be observed. As a check on the ``realism" of each model, we determine whether or not the disks present in these models are stable. Our basis for disk stability is the Toomre $Q$ parameter, defined as:
\begin{equation}
    Q = \frac{c_{\rm s}\kappa}{\pi G \Sigma}
    \label{eq:toomreq}
\end{equation}
for a disk with sound speed $c_{\rm s}$, epicyclic frequency $\kappa$, and surface density $\Sigma$. We adopt the smallest value of $Q$ in the midplane of each model as a lower bound on stability. 

\begin{figure}
    \centering
    \includegraphics[width=0.95\textwidth]{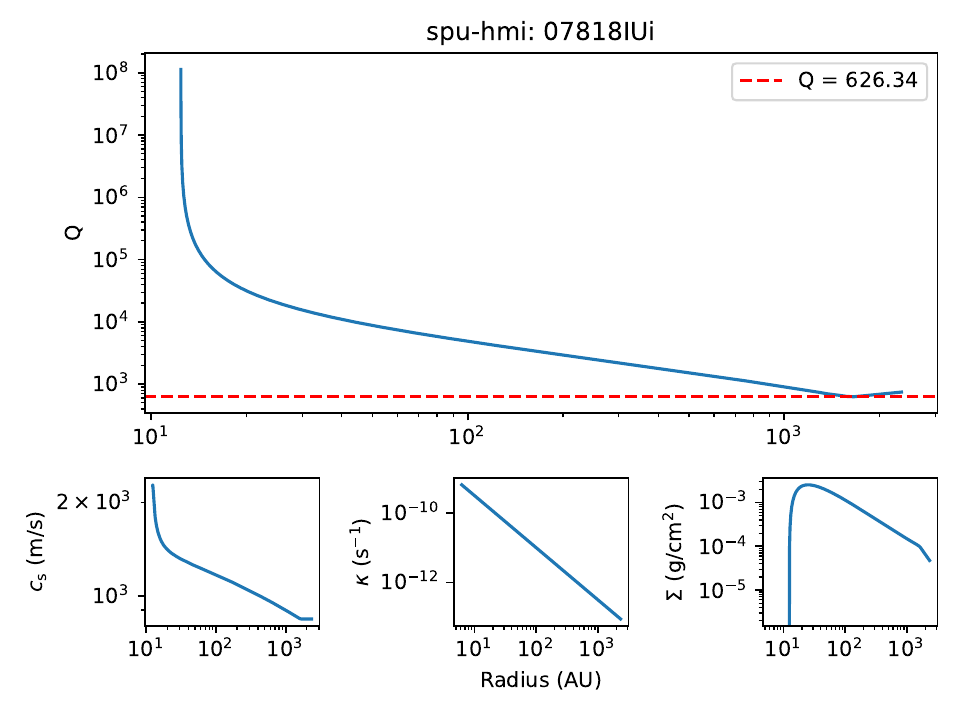}
    \caption{$Q(R)$ for a disk in one of the R17 models. The floor for $Q$ is plotted in red; we use this value to determine the minimum stability of a disk against collapse. \textit{Lower row:} Our calculated sound speed $c_{\rm s}$ \textit{(left)}, epicyclic frequency $\kappa$ \textit{(middle)}, and surface density $\Sigma$ \textit{(right)} in the disk as a function of $R$.}
    \label{fig:q}
\end{figure}

To calculate $Q$ as a function of radius $R$ for each disk, we find the variables in \eqref{eq:toomreq} as functions of $R$ individually. Assuming gas in the disk is roughly ideal allows us to calculate the sound speed using the following:
\begin{equation}
    c_{\rm s} = \sqrt{\frac{P}{\rho}} = \sqrt{\frac{k_{\rm B}T(R)}{\mu m_{\rm p}}}
    \label{eq:soundspeed}
\end{equation}
where we can calculate the mass-weighted average temperature $T(R)$ directly from the models, as described in Section \ref{sec:3.1}. We adopt a $\mu$ of 2.4 in accordance with the calculated mean molecular weight per free particle from \citet{kauffmann2008}. To calculate $\Sigma(R)$, we begin at $M_{\rm disk}(R)$, which we have for each disk directly from the dust density profiles. We average the mass over a circle with radius $R$ to calculate $\Sigma$. For the epicyclic frequency, we assume all disks are Keplerian. $\kappa$ is therefore equivalent to the angular frequency of the disk $\Omega(R)=\sqrt{GM(R)/R^{3}}$. However, R17 does not directly assign masses to the central sources in its YSOs, which are included alongside $M_{\rm disk}(R)$ in the total mass profile $M(R)$. We set a floor of 0.1 $M_\odot$ for each source to calculate a minimal $\kappa$ for each model. Based on Eqn. \eqref{eq:toomreq}, a lower bound on $\kappa$ will give the minimum $Q$ possible for an associated disk, $Q_{\rm min}$. Since there are no assigned stellar masses for the central sources, a lower bound on $Q$ provides a measure of the lowest possible level of stability for a disk (i.e. if a disk is stable by this metric, it is virtually certain to actually be stable.) The actual $Q$ for any model may then be revised upwards based on actual masses assigned to the central sources. We choose a minimum source mass roughly at the substellar boundary, given that a brown dwarf source will not reach the minimum temperature of 2000 K covered by R17. A visualization of how we calculate $Q(R)$, and therefore $Q_{\rm min}$, can be seen in Figure \ref{fig:q}.

\subsection{Extinction}\label{sec:3.4}
We have calculated the circumstellar $A_{\rm V}$ for each model along the sightlines where the SEDs are defined. This is a property that was included in the R06 model grid, but was not present in R17.  Since this is a quantity that provides additional information about the physical state of a YSO, useful when using these models as templates for SED fitting, we include it in our release.

$A_{\rm V}$ along a line of sight is equivalent to $1.086\times$ the sightline's optical depth $\tau$. We can calculate the extinction as a function of frequency $A_\nu$ using:
\begin{equation}
    A_\nu = 1.086\tau = 1.086\int \kappa_{\rm \nu; dust}\times\rho_{\rm dust}\times ds
    \label{eq:av}
\end{equation}
where we have the dust density present in every cell from Section \ref{sec:3.1} and the opacity of the dust as a function of frequency, $\kappa_\nu$, from Section \ref{sec:2.2}. We calculate $A_{\rm V}$ by evaluating $A_\nu$ at the standard V-band wavelength of 551 nm. As in previous sections, we only consider dust above the ambient density.

\newpage
\subsection{Convenience additions}\label{sec:3.5}
In this section, we outline additions we have made to the models that do not require a significant investment in time or resources to replicate, but still provide new information or functionality to the models as presented in R17.

The largest such addition is the introduction of several new sets of convolutions of the model SEDs. At release, the SEDs in R17 were convolved with filters from instruments across the visible and IR. Since the release of R17, however, new IR instruments have been introduced. The SEDs also span a wavelength range that makes them useful for work outside the near-IR. We have convolved the SEDs with filters from JWST, Paranal, additional filters from Herschel, and ALMA bands 3 and 6. Most of these filters were obtained from the SVO's Filter Profile Service \citep{SVO2012,SVO2020}. Beyond that, we have created a script to easily integrate convolutions either from the SVO or user-defined filters into the existing infrastructure, included with the dataset released alongside this paper.

We have also calculated the infrared spectral index $\alpha$ for all SEDs within each aperture, defined by the following equation:
\begin{equation}
    \alpha = \frac{d\, {\rm log} \left(\lambda F_\lambda\right)}{d\, {\rm log}\lambda}
\end{equation}
This quantity can be used to place observed YSOs into particular observational classes, which is a common tactic used to gain physical insight into the evolutionary state of a YSO (we discuss this practice more in Section \ref{sec:4.2}). We take $\alpha$ to be the slope of the log-space line joining the flux at 2 and 25 microns, the endpoints of the standard wavelength range \citep{kennicutt2012}. We consider only the endpoints when calculating the spectral index, i.e. we do not fit to all points within the wavelength range to determine the slope.

\begin{figure}
    \centering
    \includegraphics[width=0.99\textwidth]{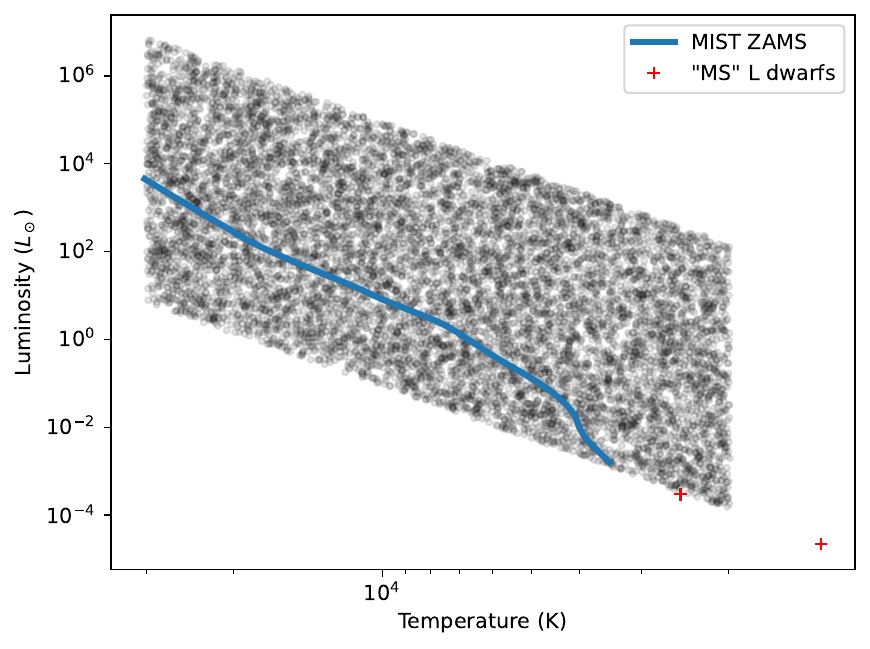}
    \caption{Main sequence taken from MIST evolutionary tracks, plotted over the central sources of \texttt{s---s-i}, a geometry that only includes bare stars. The evolutionary tracks used to construct the main sequence assume $v/v_{\rm crit}$ = 0.4 and [Fe/H] = -2.00. The individual data points are rough estimates of the position of ``main sequence" L dwarf objects across their temperature range using values from \citet{kirkpatrick2005}, plotted to give an indication of how the main sequence continues below temperatures accessible by our chosen MIST tracks.}
    \label{fig:ms}
\end{figure}

In addition, we have made explicit the luminosity of the central source of each model. As an example of its utility, we evaluate the extent to which random sampling of stellar radius and temperature for each model results in sub-main-sequence sources. For our main sequence, we use the ZAMS (EEP \# = 2) temperatures and luminosities from a set of MIST evolutionary tracks \citep{dotter2016,choi2016,paxton2011,paxton2013,paxton2015}. We adopt the values from the tracks with $v/v_{\rm crit}$ = 0.4 and [Fe/H] = -2.00; we choose this metal abundance to correspond with the simulated stellar photospheres of \citet{brott2005} and \citet{castelli2003} used in the models (see R17 section 3.2.1 for more detail). In Figure \ref{fig:ms}, we plot this main sequence over the region of T-L space covered by the models.
We use these main-sequence $T$ and $L$ values to interpolate the main-sequence luminosity $L_{\rm MS}$ of our models based on their temperatures. We find that 23\% of all models have $L$ $<$ $L_{\rm MS}$ and therefore occur below the main sequence. Another 24\% of all models have temperatures that are too low to be interpolated. As a check on the placement of these sources relative to the main sequence, we use the temperature range and average radius of L dwarf objects from \citet{kirkpatrick2005} to map out a rough continuation of the main sequence below MIST values. Based on this extrapolation, most--if not all--of these cooler sources would likely occur above the main sequence. We include the MIST $T_{\rm MS}$/$L_{\rm MS}$ values in our data release for those interested in repeating this analysis. In Appendix \ref{ap:lum}, we also use this luminosity information to check the energy conservation of the radiative transfer performed by \texttt{Hyperion}.

Finally, we have included the radii where non-ambient density structures begin and end for each YSO model in units of physical distance. Previously, the outer radii for envelopes were not included, and the inner radii for every model were either in terms of the variable quantity $R_{\rm sub}$ or not included.

\subsection{Completed YSO models}\label{sec:3.6}
\begin{figure}
    \centering
    \includegraphics[width=0.7\textwidth]{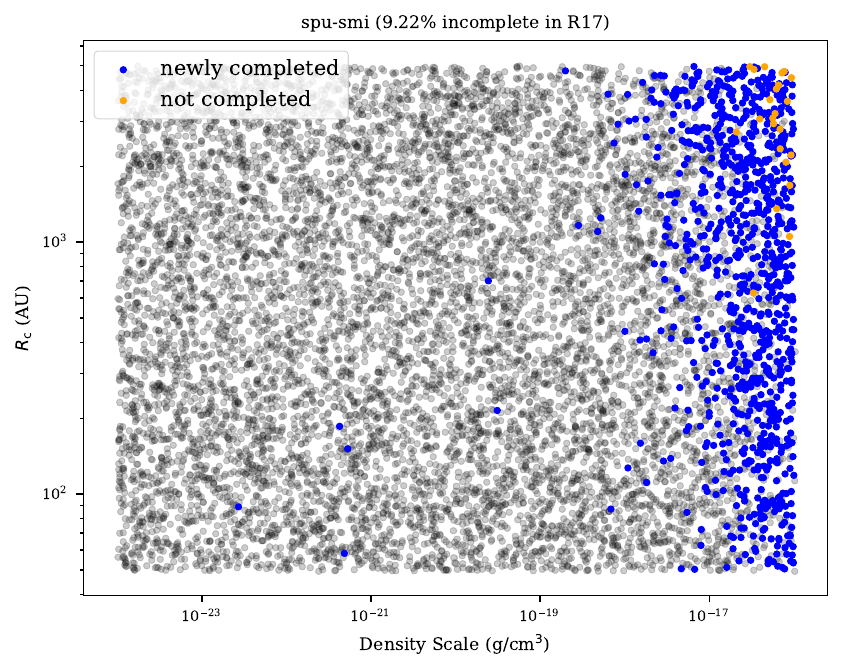}
    \caption{A 2D projection of the R17 model parameter space for a geometry with a U76 density profile \eqref{eq:ulrich}. The free parameters are the centrifugal radius $R_{\rm c}$ and density scale $\rho_{0}$. Highlighted models were not run to completion in \texttt{Hyperion} in the initial release; those in blue have since been completed. There is a bias towards higher-density envelopes resulting from high optical depth. The apparent ``wedge" is caused by the radial behavior of U76 density profiles. All instances of radial dependence are scaled as $R_{\rm c}/r$, causing the density to fall more slowly as $R_{\rm c}$ increases, particularly when $r < R_{\rm c}$. As a result, the optical depth of envelopes is correlated with increasing $R_{\rm c}$. The net effect is to depopulate the high-density models.}
    \label{fig:missing}
\end{figure}

Some models included in the initial release of R17 had assigned parameter values but did not have associated SEDs. These models were not run successfully run to completion in \texttt{Hyperion}. While multiple geometries had incomplete models, those with U76 density profiles exhibited a lower rate of completion on average. (The exact completion rate varies with geometry features, and tends to be lower for the U76 models without cavities.) The incomplete models exhibit a bias towards denser envelopes, as seen in Figure \ref{fig:missing}; the failure to complete is therefore likely due to the high optical depth resulting from the shallow fall-off of U76 density profiles with radius. The models are otherwise distributed akin to the broader set.

We have run incomplete models in \texttt{Hyperion}, version 0.9.10, to further fill in the parameter space. The models run for the initial release were performed on a cluster of 120 cores; we use at minimum a cluster of 256 cores. This is sufficient to complete the majority of outstanding models, though some particularly intractable models require $\geq$1500 cores to complete within the allowable time. These holdouts are so optically thick that there may not be a sufficient number of photons to construct an SED, particularly at long wavelengths. As a result, the impact of their exclusion from the model SEDs is less pronounced than it might be otherwise; however, we do plan to run these models for the sake of completeness.

In the initial release, model runs produced a raw output file that would then be post-processed according to Section 4.2.4 of R17. We have followed the same procedure for our newly completed models and will continue to do so for future additions, both of the as-yet-uncompleted models from the initial release and any newly created models we add.

\subsection{Content updates}\label{sec:3.7}
In the course of this work, we have primarily built on top of the existing R17 models without making changes. In using our additions to attain further results, though, we have come across a property of the model SEDs that we feel is worth altering in our release. 

In the data products published alongside R17, each model's associated SED is defined in a common set of apertures, as described in Section \ref{sec:2.3}. However, each SED is calculated within its own set of apertures based on the physical size of its associated model. This calculated SED is then interpolated to the final set of apertures. For models with large inner radii, the smallest calculation aperture may be larger than apertures in the final set. In these cases, \texttt{Hyperion} does not have sufficient resolution to capture the SED inside these smaller apertures. In the initial release, such cases were handled by assigning any out-of-bounds apertures the same flux as in the closest in-bounds aperture. This resulted in inner apertures in the final set appearing brighter than they should, given the amount of material they contain. In our release, we have elected to instead assign NaN values to model SEDs in any aperture below the resolution of the model, so that these results cannot introduce noise.

It is also possible for the final set of apertures to extend outside of those native to a model (into an ambient medium, should the model have one.) However, since flux from material at the background density and temperature is subtracted in post-processing, any such additional material does not produce any additional flux. Assigning these apertures the closest in-bounds value is therefore self-consistent and allows these models to retain their use as templates for SED fitting, so we do not alter this component of post-processing. 

In our updated set of models, every SED has been appropriately interpolated into the final set of apertures. However, this set of apertures places its own limit on the resolution of the SEDs. Care should therefore be taken when applying these models to high-resolution data. As an example, JWST's maximum resolution of 0.1" has a physical size $<$ 100 AU for distances within 1 kpc, so the models would fail to predict JWST fluxes for nearby targets and cannot be used to fit data at that resolution.

\section{Results}\label{sec:4}
In this section, we present use cases for our updated version of the R17 models that are made possible through our additions.

\subsection{Mass measurements}\label{sec:4.1}
A common assumption made when inferring the mass of a pre- or protostellar core from an observed flux is that the dust in the core is optically thin. Low optical depth permits the calculation of mass through the following equation:
\begin{equation}
    M_{\rm core} = \frac{S_{\rm\nu,dust}d^{2}}{\kappa_{\rm\nu,dust\& gas}B_{\nu}(T_{\rm dust})}
    \label{eq:optthinmass}
\end{equation}
for an observed flux $S_{\rm\nu,dust}$ at a distance $d$ with total material opacity $\kappa_{\rm\nu,dust \& gas}$ and dust temperature $T_{\rm dust}$. In addition, a constant dust temperature (often $\approx$20 K) is generally adopted. The accuracy of mass measurements made using Eqn. \eqref{eq:optthinmass}--and any further uses of those measurements--hinges on those assumptions. With our newly calculated masses, we test the accuracy of the optically thin, isothermal assumption.

We compare the amount of mass inferred using Eqn. \eqref{eq:optthinmass} and the flux from a model to the mass we calculate is present in that same model. We adopt the same dust properties used in R17, as described in Section \ref{sec:2.1}. We scale the opacities from D03, plotted in Figure \ref{fig:opacities}, to include gas mass at our assumed GDR. Our method for mass calculation is described in Section \ref{sec:3.1}. Since we are interested in evaluating the impact of the assumed dust temperature, we consider both a constant 20 K and a mass-weighted average temperature within each aperture, which we calculate as described in Section \ref{sec:3.2}. We consider the mass-weighted average temperature on the expectation that a modified blackbody at this temperature will reasonably approximate the total emission of dust in an optically thin system. We use our calculated ``line-of-sight" masses and temperatures, which are by construction observational analogues.

To begin, we examine a specific mock-observational case: a flux of $10$ mJy at 1.1 mm at a distance of 1 kpc within an aperture of $\approx 1000$ AU. We consider models over all geometries with envelopes that exhibit a flux of $10\pm 1$ mJy in that aperture. We make additional cuts to the set of models that also have disks, keeping those with an average optical depth $\bar{\tau}_{\rm disk}<10^{-3}$ and $Q_{\rm disk} > 0.1$. We calculate $\bar{\tau}_{\rm disk}$ by multiplying the dust opacity at 1.1 mm by the average disk surface density. The resulting models are those that best follow the assumption of low optical depth. They are therefore best positioned to illustrate the relationship between the mass we calculate to be present (our ``true" mass) and the mass inferred through Eqn. \eqref{eq:optthinmass}.

\begin{figure}
    \centering
    \includegraphics[width=0.75\textwidth]{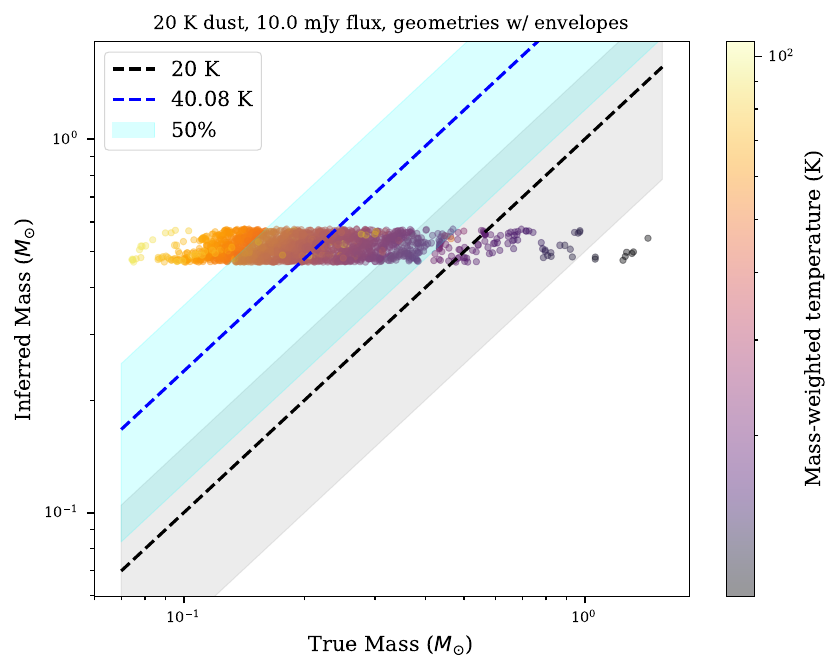}
    \includegraphics[width=0.49\textwidth]{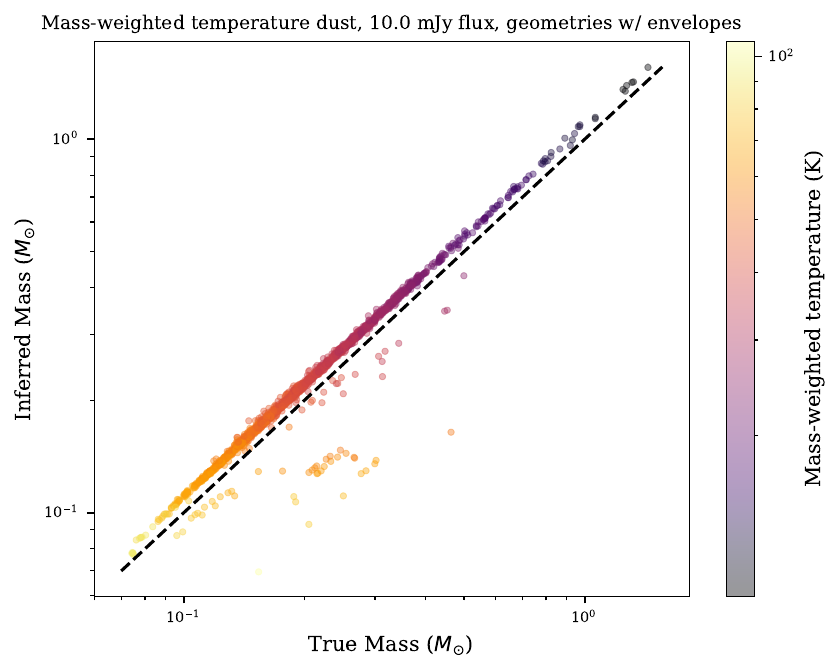}
    \includegraphics[width=0.49\textwidth]{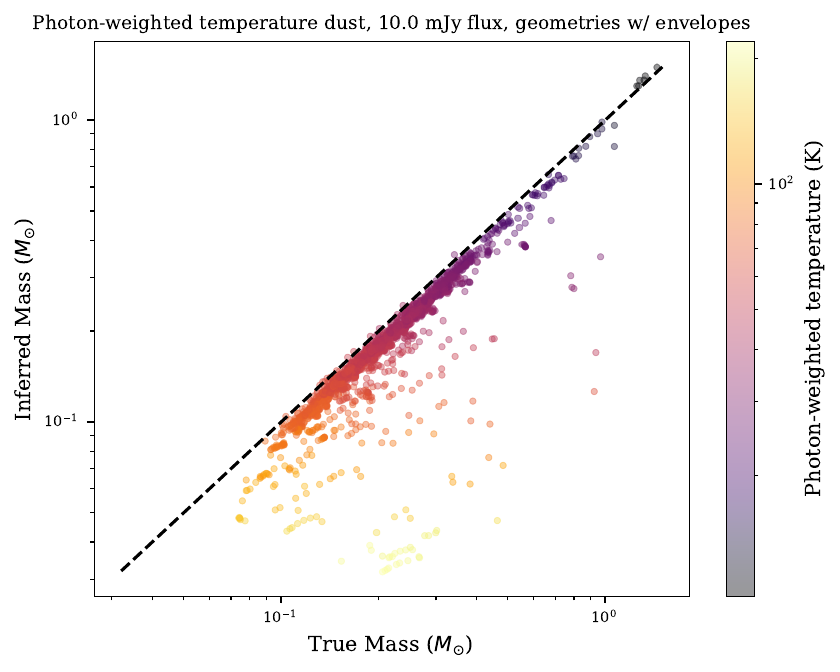}
    \caption{\textit{Top:} Mass inferred for a set of models assuming a constant 20 K dust temperature plotted against our ``true" masses for each model, calculated along the line of sight. \textit{Bottom:} Mass inferred using the mass-weighted (\textit{left}) and mass$\times$temperature-weighted/``photon-weighted" (\textit{right}) average dust temperature of each model plotted against its calculated mass. Each model exhibits a 10$\pm$1 mJy flux at 1 millimeter in an aperture of $\approx$1000 AU at a distance of 1 kpc. All model geometries with envelopes are included in this plot. Models included in this plot are optically thin, based on our criteria in \S\ref{sec:4.1}. The expected scenario for optically thin dust, $y=x$, has been plotted for comparison \textit{(black dashed line)}. In the constant-temperature case, another line is plotted where models at the median mass-weighted dust temperature, $\approx$40 K, occur. The shaded regions in the top plot indicate where models have an inferred mass within 50\% of the mass corresponding to their respective dust temperatures. Models are colored by their average dust temperature; models with dust at a higher temperature that emits more brightly tend to be less massive than models with cooler dust emitting the same amount of flux within the same aperture, as expected.}
    \label{fig:masscomps}
\end{figure}

Results from the comparison are plotted in Figure \ref{fig:masscomps}. When assuming a constant dust temperature of 20 K (the top plot in Figure \ref{fig:masscomps}) there is no apparent correlation between the inferred masses and true masses. This indicates that even when observing YSOs that satisfy the $\tau \ll 1$ criterion, assuming isothermal dust at 20 K will not reliably provide accurate mass measurements. Under these assumptions, the majority of models have their mass overestimated by a factor of 2 or more.

Conversely, assuming the dust in each model is at its weighted-average dust temperature causes consolidation into a linear relationship between inferred mass and true mass, as should be the case if Eqn. \eqref{eq:optthinmass} holds. We consider two weighting schemes; one by the mass present in each cell, as calculated in Section \ref{sec:3.2}, and one by the product of the mass and dust temperature in each cell as an approximation of the photon-weighted dust temperature at long wavelengths. These are the bottom left and bottom right plots in Figure \ref{fig:masscomps}, respectively. In both cases, we see good agreement between the inferred and true masses for our sample of optically thin YSOs. In the mass-weighted case, there is a strong correlation between the inferred and true masses, although the inferred masses are systematically greater than the true masses by approximately 10\%. In the photon-weighted case, the inferred mass effectively serves as a lower limit to the true mass, and 46\% of models are within 10\% of a one-to-one correlation. (87\% of models are within 20\%.)

\begin{figure}
    \centering
    \includegraphics[width=0.8\textwidth]{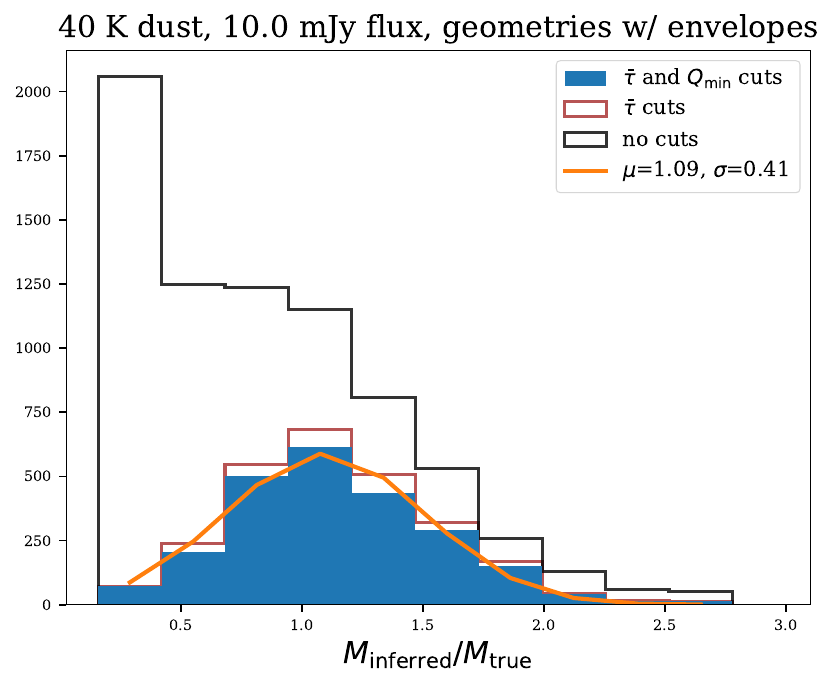}
    \caption{Histograms of the ratio of the inferred mass to the true mass of models in the observational case in Figure \ref{fig:masscomps}, assuming dust at the median mass-weighted average dust temperature of 40 K. \textit{Filled:} With cuts made based on the average optical depth and stability of disks (\S\ref{sec:4.1}). A Gaussian is fit to the histogram as a model for the peak and width of this distribution. The majority of the models fall within a range of .4 around the mean of 1.1, which translates to the majority of models having an inferred mass of $\approx$70-150\% of our calculated value. \textit{Unfilled, brown:} With cuts made based only on the average disk optical depth. While this histogram includes models that are unlikely to be observed based on their stability (i.e. small $Q_{\rm min}$), the disk stability cut has a minimal effect on the results. \textit{Unfilled, black:} No cuts made at all. The shape is no longer Gaussian, as this includes models that are not optically thin and therefore have more mass than is visible from the observed flux. Including models that exhibit a flux without making cuts based on optical depth decreases the median dust temperature, as additional cold material in disks is hidden from view by higher optical depth.}
    \label{fig:masshist}
\end{figure}

It is clear that an assumption of 20 K dust is not representative of the models in this specific case. However, if we instead assume that all dust is at the median mass-weighted average dust temperature of our set of optically thin models with $S_{\rm 1.1 \; mm,\; 1000 \; AU} = 10\pm1$ mJy, the picture improves. In our test case, the median temperature is approximately 40 K. Returning to Figure \ref{fig:masscomps}, we have also plotted a line showing where models with an average dust temperature of 40 K fall. Per Figure \ref{fig:masshist}, more than half of the models in our sample would have inferred masses within $\approx$50\% of their true mass if we assume 40 K dust. These results suggest that the assumption of a constant dust temperature to measure a core mass may be appropriate as long as the dust is at a sufficiently low optical depth and the temperature is representative of the dust in the core. However, if the dust is optically thick but assumed to be optically thin, the mass will be severely systematically underestimated. 

\begin{figure}
    \centering
    \includegraphics[width=0.9\textwidth]{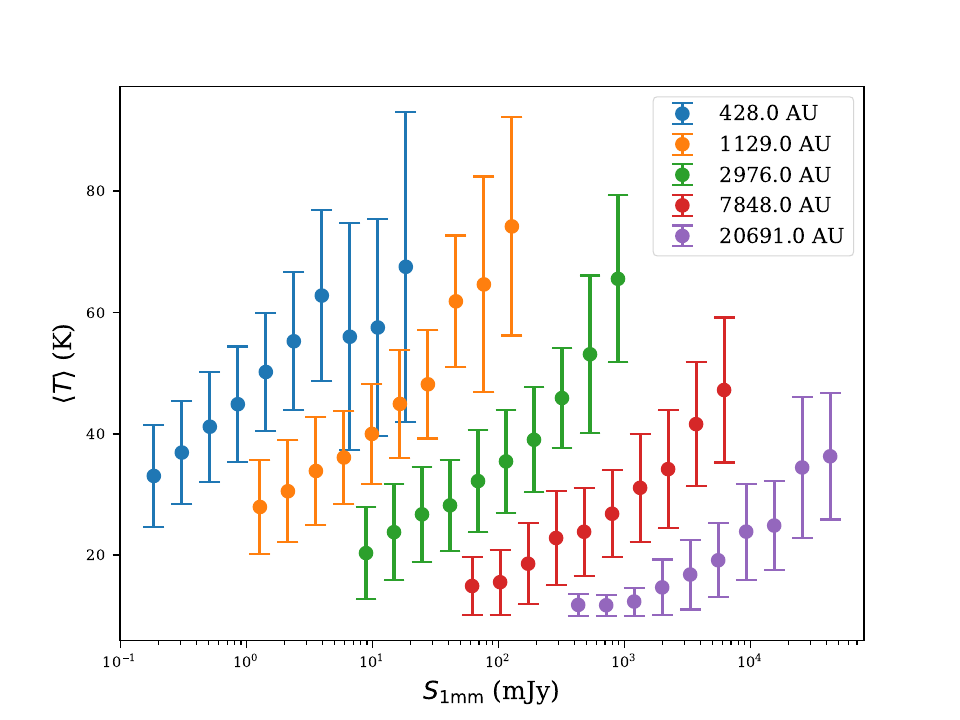}
    \caption{The median mass-weighted average dust temperature of models exhibiting a particular flux. All models used to derive these statistics are optically thin, based on our criteria in \S\ref{sec:4.1}. As the flux observed from a YSO increases within the same aperture, the temperature of the models that represent it tends to increase as well. Error bars are the median absolute deviation of the dust temperatures. Temperatures in the 428 AU aperture exhibit non-monotonic behavior with increasing flux, unlike in larger apertures. This behavior is endemic to large fluxes within smaller apertures, so care should be taken when applying these results to bright objects observed with small beams.
    }
    \label{fig:temp_v_flux}
\end{figure}

To provide a way to systematically choose a more correct dust temperature than the canonical 20 K, we repeat the procedure of identifying models that exhibit a particular flux within a given aperture across a wider range of fluxes and aperture sizes. For each resulting population, we find its median mass-weighted average dust temperature. We compare the masses inferred using these temperatures to the true masses. As in our test case, the majority of inferred masses are within 50\% of the true masses across all apertures as long as an appropriate temperature is assumed. In Figure \ref{fig:temp_v_flux}, we plot the median of the mass-weighted average dust temperatures of the models that exhibit a particular flux. Within a single aperture, the average dust temperature of the models tends to increase with the observed flux. Our result provides the most appropriate dust temperature to use for mass measurements based on the measured flux and aperture size. 

The temperature floor and physical parameter sampling limit the range of masses to which these temperature assumptions apply. By construction, the models in R17 have a floor dust temperature of 10 K. Consequently, we are unable to extend our analysis into temperatures below 10 K. This imposes an unphysical lower limit on dust temperature as a function of flux in Figure \ref{fig:temp_v_flux}. We expect this limitation to be more pronounced as the flux measured in an aperture decreases. 

We are also limited in our analysis by a dearth of models at higher fluxes within each aperture. As an example, the number of models that exhibit a flux of $S_{\rm 1000 \; AU} > 100$ mJy is insufficient to repeat Figure \ref{fig:masshist}. We therefore have less confidence in the expected dust temperatures in this area of flux space. However, we expect that while the brighter end of flux space in each aperture is poorly sampled, models in this space are less likely to occur in nature or are otherwise not relevant for our purposes here. Producing a higher flux from dust while remaining within the same aperture requires some combination of a higher temperature and more dust mass. Higher temperatures require increased energy input, which in turn requires more luminous central sources. These are less common than dimmer sources in the models, which is also generally true for observed stars. The extent to which this flux space is depopulated in the models by the dearth of extremely bright sources is therefore likely mirrored in nature. Meanwhile, introducing more dust to the same physical space will increase its optical depth. Past a certain amount of additional mass, then, a model will likely fail to meet the criterion of low optical depth for inclusion in the sample.

As an addition to the previous caveat, we have low confidence in our predicted temperatures within apertures smaller than 1000 AU for large fluxes. To illustrate, the median dust temperature in the smallest aperture in Figure \ref{fig:temp_v_flux} does not uniformly increase alongside the measured flux above $\approx 5$ mJy, which runs counter to our expectations. There is a known decrease in sample size at higher in-aperture fluxes, and smaller apertures are also less likely to have a long-wavelength flux due to post-processing for noise (see Section 4.2.4 of R17.) Taken together, and with a lack of apparent correlation to any parameter of the models, these facts imply that this non-monotonic behavior is related to a sample size issue. Our suggestion is to treat predicted temperatures at high fluxes within small apertures with care, as they may be artificially lowered. Mass estimates made using these temperatures, then, are effectively upper bounds on the true mass.

\subsection{YSO classification}\label{sec:4.2}
A common tactic used to constrain the physical state of a YSO via observation is to determine its SED class, based on its near- to mid-infrared spectral index. These observational classes are thought to correspond to related evolutionary stages. For example, YSOs of Class I have positive spectral indices, which implies that the SED is dust-dominated. In turn, dust domination of the SED is thought to occur during the stage when most of the mass of a YSO is in its envelope \citep[][D14]{dunham2014}.

The position of a YSO in color space is often used to determine the class \citep[e.g.][G09]{gutermuth2009}. However, since class is a fundamentally observational method of characterization, there is no guarantee that it actually corresponds to the physical stage. YSOs may be mischaracterized due to observational effects such as inclination or reddening. R06 attempted to provide a more directly physically motivated tool for identifying and characterizing YSOs by placing its YSO models in color space grouped by physical stage as opposed to class. Our additions to R17 allow us to extend this analysis by comparing the positions of YSOs grouped by class and by stage in color space. We can use these groups to determine the extent to which different classes and stages can be distinguished observationally, and more broadly, the extent to which the concepts of ``class" and ``stage" are related.

In each diagram, we include an arrow that indicates the effect of ten magnitudes of V-band extinction ($A_{\rm V}=10$). We assume extinction takes place according to the \citet{fitzpatrick1999} extinction law as modified by \cite{indebetouw2005}. Throughout this section, diagrams are colored according to the density of models in color space, which is done individually for each class and stage. In Appendix \ref{ap:ccds} we present an alternate view of these diagrams that instead show where in color space each class and stage is most dominant.

\subsubsection{Class}\label{sec:4.2.1}
\begin{figure}
    \centering
    \includegraphics[width=0.99\textwidth]{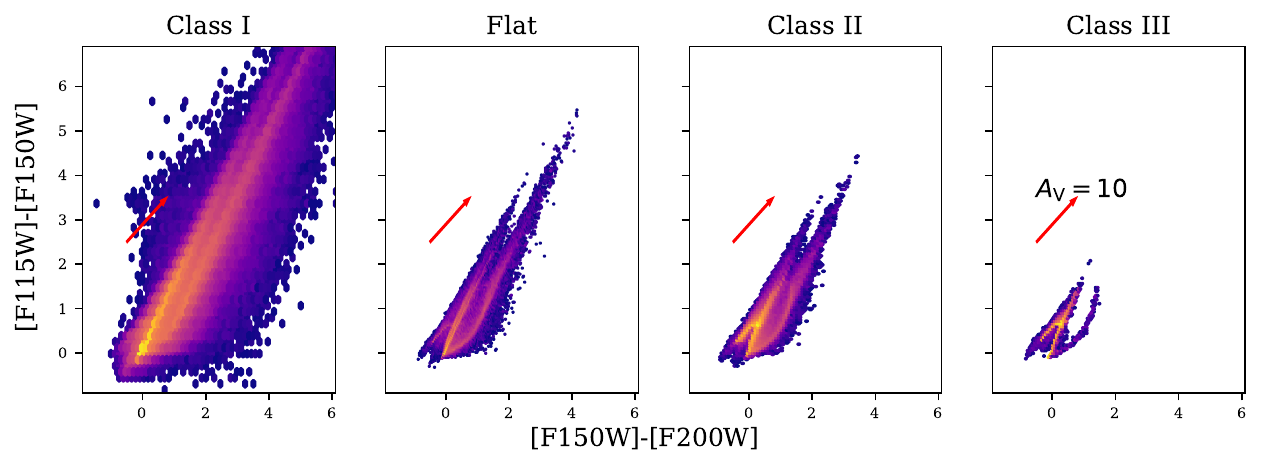}
    \includegraphics[width=0.99\textwidth]{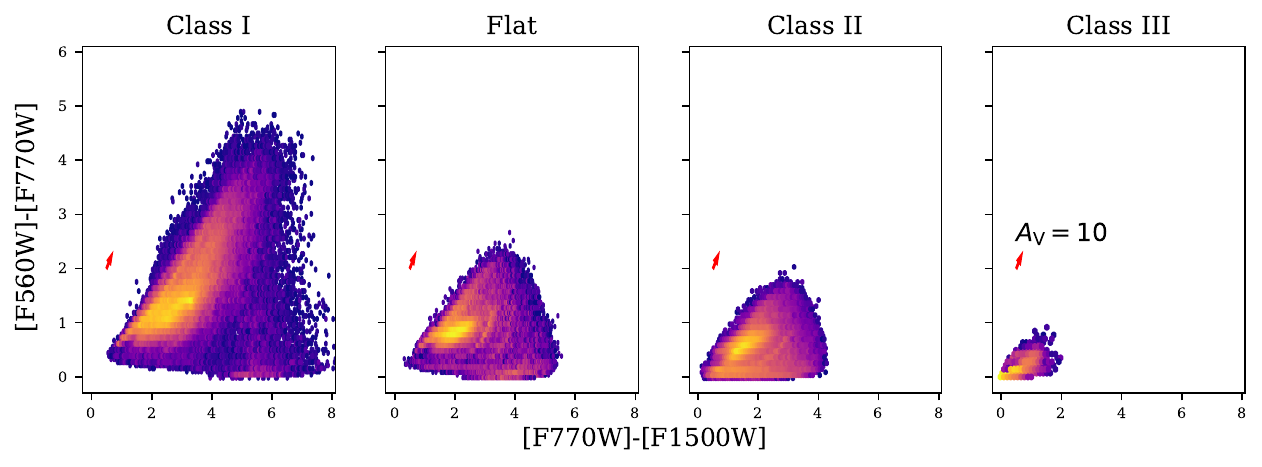}
    \includegraphics[width=0.99\textwidth]{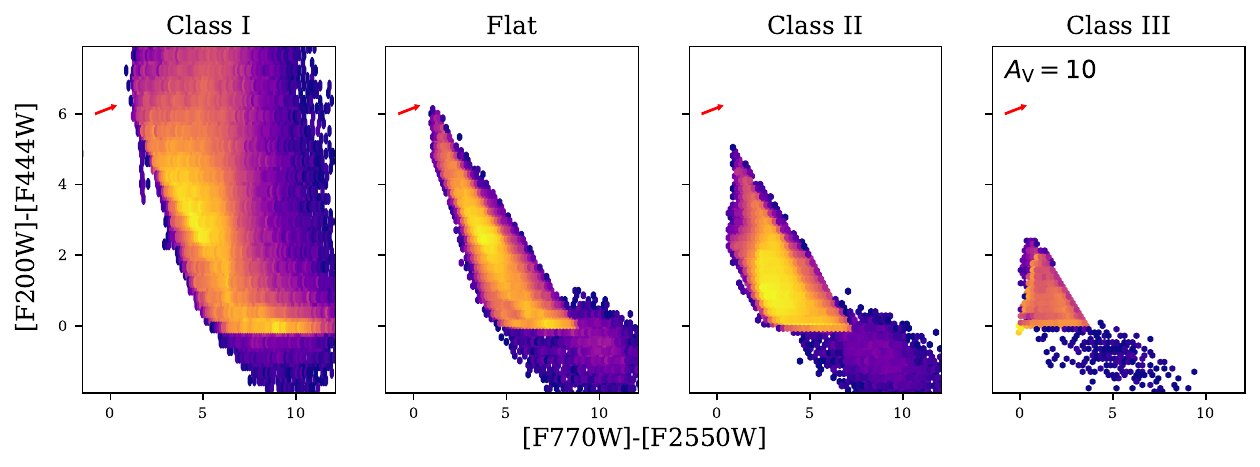}
    \caption{Color-color diagrams of models with different classes, based on the classification scheme from D14. We provide one in NIRCam filters (\textit{top}), one in MIRI filters (\textit{middle}), and one spanning the full wavelength range accessible by each instrument (\textit{bottom}). An arrow is plotted on each panel showing the effect of 10 magnitudes of visual extinction according to the \citet{fitzpatrick1999} extinction law (as modified by \citet{indebetouw2005}). All colors are based on model SEDs calculated within a 1000 AU aperture at a distance of 1 kpc.}
    \label{fig:classccds}
\end{figure}

Classification of an SED is based on its spectral index, which we have calculated (\S\ref{sec:3.5}). Throughout this section, we adopt the spectral classification scheme from \citet{greene1994} as described in D14, which is as follows:
\begin{itemize}
    \item Class I: $\alpha \geq 0.3$
    \item Flat: $-0.3 \leq \alpha < 0.3$
    \item Class II: $-1.6 \leq \alpha < -0.3$
    \item Class III: $\alpha < -1.6$
\end{itemize}
D14 also includes a ``Class 0" from \citet{andre1993} for protostars that are undetectable in NIR wavelengths but are identified via millimeter continuum detection of dust. Observationally, a source is Class 0 if it has $L_{\rm smm}/L_{\rm bol}$ $>$ $0.5\%$, where $L_{\rm smm}$ covers $\lambda$ $\geq$ 350 $\mu$m. We assign Class 0 to all models that meet this criterion and do not already have a defined spectral class. The remaining SEDs (approximately 5\% of the total number of SEDs) are not assigned a spectral class; these cover both SEDs which fail to meet the criteria for all other classes as well as those that are as yet incomplete (see Section \ref{sec:3.6} for details). In Figure \ref{fig:classccds}, we separate the models by class and visualize them in JWST color space. (We do not include Class 0 models, as by definition they will generally not be detectable by JWST.)

In NIRCam color space (Figure \ref{fig:classccds}, top) there is a general trend towards the blue from Class I to Class III. Class I models span the largest range in both colors and are capable of having [F115W]-[F150W] $>$ 5 and [F150W]-[F200W] $>$ 4, redder than other classes can achieve. Flat and Class II YSOs generally occupy the same region of color space; most have [F115W]-[F150W] $<$ 3 and [F150W]-[F200W] $<$ 2, with a small fraction of redder models. Class III YSOs mostly have both colors $<$ 2. Class I YSOs occupy a unique area of color space, so it is theoretically possible to distinguish them from the other classes. However, this area of space is located along the direction of extinction, so significant degeneracy between interstellar and circumstellar extinction is expected for more deeply-embedded and/or more distant sources. Proper classification therefore hinges on being able to deredden appropriately.

In MIRI (Figure \ref{fig:classccds}, middle), as for NIRCam, there is a clear trend blueward with increasing class number. Class I models are bounded by [F560W]-[F770W] $<$ 5 and [F770W]-[F1500W] $<$ 8. Flat models can have colors out to 3 and 5, respectively, while Class II models remain within 2 and 4. Class III models are generally bounded by 1 and 2. Unlike NIRCam, there are distinctions in this color space that are separable from interstellar extinction, which affects [F560W]-[F770W] more strongly than [F770W]-[F1500W]. Models with [F770W]-[F1500W] $>$ 5, caused by warm disks with significant emission around 15 microns, are probably Class I. YSOs are also progressively more likely to have zero color along either axis, but particularly [F560W]-[F770W], going from Class I to III.

In a combined NIRCam + MIRI color space (Figure \ref{fig:classccds}, bottom), different classes are easily separable. Despite similar ranges in NIRCam colors (as above), the range in [F770W]-[F2550W] shifts between Classes I, II, and III such that they occupy different ``slices" of color space. In this color space, the effect of extinction is most pronounced in [F770W]-[F2550W], meaning that accurate dereddening is important to be able to distinguish between sources with [F200W]-[F444W] close to 0. On the whole, however, degeneracy with extinction can be avoided by combining near-IR and mid-IR photometry.

\subsubsection{Stage}\label{sec:4.2.2}
\begin{figure}
    \centering
    \includegraphics[width=0.99\textwidth]{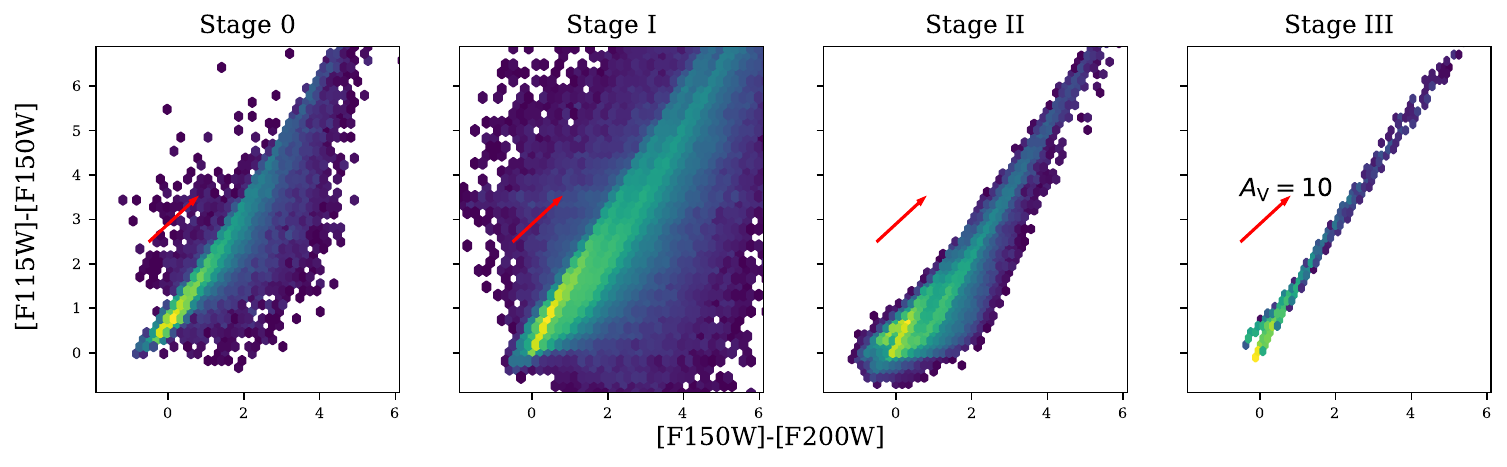}
    \includegraphics[width=0.99\textwidth]{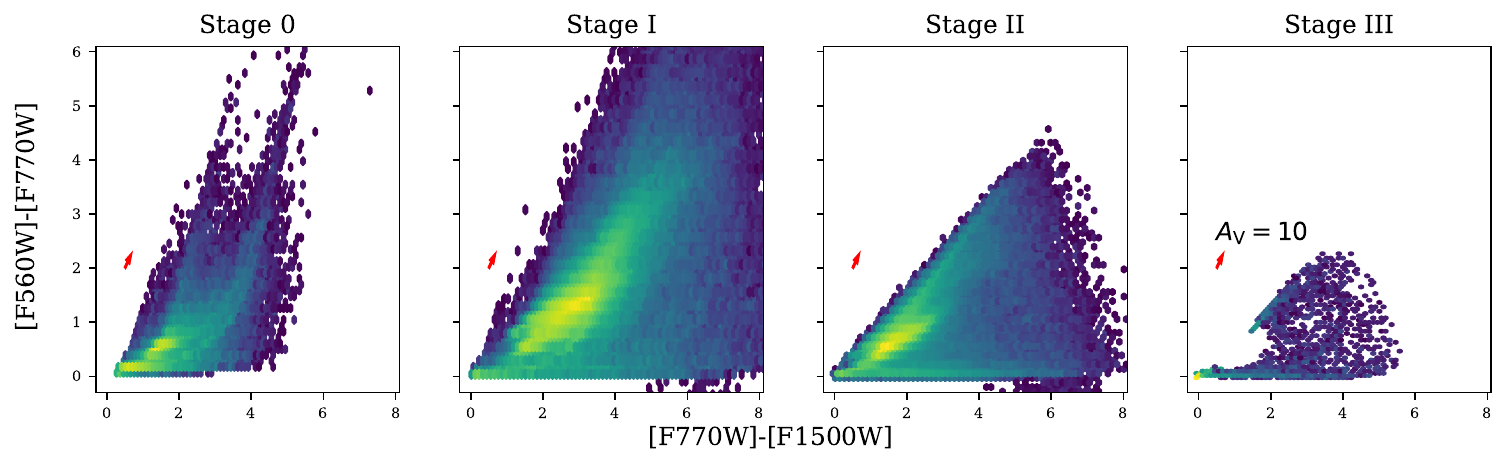}
    \includegraphics[width=0.99\textwidth]{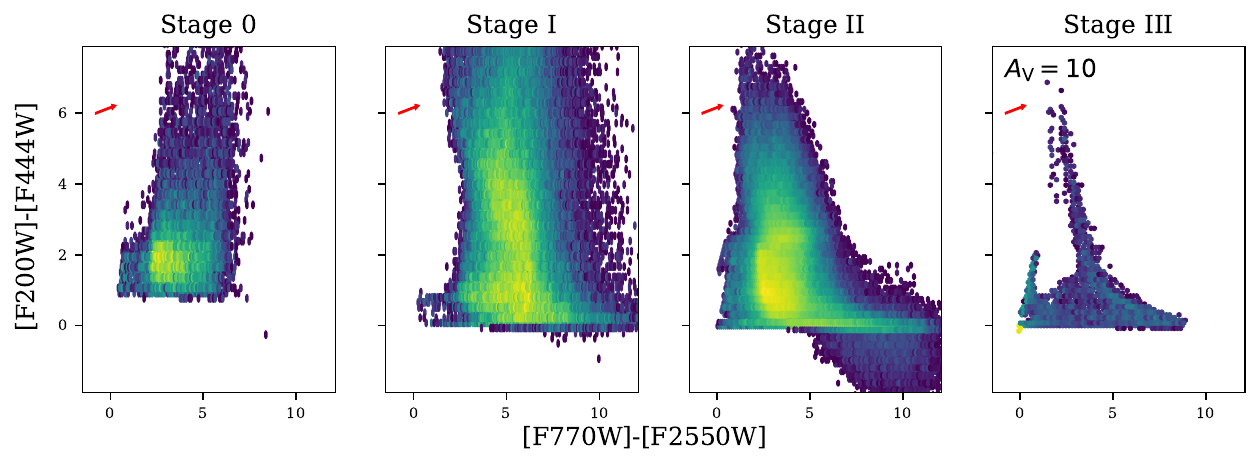}
    \caption{The same as Figure \ref{fig:classccds}, but with models separated by physical stage instead (see \S\ref{sec:4.2.2} for details).}
    \label{fig:stageccds}
\end{figure}

We adopt the following ``stagification" scheme, which uses the definitions from \citet{crapsi2008} and \citet[][E09]{evans2009} as a base:
\begin{itemize}
    \item Stage 0: $M_{\rm env}\, >\, 0.1 M_\odot$, $T_\star < 3000$K
    \item Stage I: $M_{\rm env}\, >\, 0.1 M_\odot$, $T_\star > 3000$K
    \item Stage II: $M_{\rm env}\, <\, 0.1 M_\odot$, disk present
    \item Stage III: Bare pre-main-sequence star (no envelope, no disk)
\end{itemize}
We extend the E09 definition by separating Stages 0 and I by stellar temperature on the grounds that Stage 0 sources should be too cool to be proper pre-main-sequence stars, unlike those in Stage I (i.e. Stage 0 sources will not have reached the Hayashi or Henyey tracks, which begin at approximately 3000 K). We repeat the color-color diagrams from the previous section and plot them in Figure \ref{fig:stageccds}.

Stages are difficult to distinguish in NIRCam color space (Figure \ref{fig:stageccds}, top). In each stage, models are mostly concentrated close to zero color--as dictated by the stellar photosphere models--and the models reddened by dust are largely found along the direction of extinction. The majority of all models are bounded by [F115W]-[F150W] $<$ 6 and [F150W]-[F200W] $<$ 5. Since Stage III is entirely photospheres surrounded by an ambient medium, the position of Stage III models is particularly concentrated in color space. The shape of models in earlier stages is wider around the direction of extinction by comparison. This is particularly true in Stage I, which can achieve [F115W]-[F150W] $>$ 3 at low [F150W]-[F200W] and [F150W]-[F200W] $>$ 2 at low [F115W]-[F150W]. It is therefore possible to distinguish non-Stage III YSOs from reddened stars, and Stage I YSOs in particular, using NIRCam colors. However, doing so relies on proper dereddening.

In MIRI space (Figure \ref{fig:stageccds}, middle) there is a clear evolution redward in [F770W]-[F1500W] from Stage 0 to Stage II. Models of Stages I and II reach [F770W]-[F1500W] $>$ 4, unlike the majority of Stage 0. Some of these models have [F560W]-[F770W] commensurate with the effects of extinction; however, some are also redder in [F770W]-[F1500W] compared to [F560W]-[F770W] than could be achieved by extinction alone. Stages I and II encompass the transition from envelope domination to disk domination, so the appearance of preferential reddening in [F770W]-[F1500W] for Stage I and II models shows that the mid-IR colors involving longer-wavelength emission are disk-dominated. Redder colors result from emission by heated disks, which can be seen through the small or nonexistent envelopes of late Stage I and Stage II. These longer-wavelength colors may therefore be useful in separating more envelope-dominated sources from more evolved ones that have cleared more of their surroundings. Stage III models mostly have colors largely close to zero, though as for NIRCam, some models are reddened by the ambient medium. These models tend to be redder in [F770W]-[F1500W] than expected from sightline extinction, driven by emission from a medium that is heated by higher-temperature sources.

Trends from the individual instruments are also visible in a combined NIRCam + MIRI space (Figure \ref{fig:stageccds}, bottom). Stage I models extend into NIRCam colors that are redder than Stages 0 and II ([F200W]-[F444W] $>$ 7) and Stage I and II models are generally capable of being redder in MIRI than Stage 0 ([F770W]-[F2550W] $>$ 5). This last point is especially true for Stage II, which has a higher concentration of models that are red in MIRI than the others. However, in this case, the direction of extinction is primarily redward in MIRI, and thus must be disentangled from the effects of evolution.

An examination of Figure \ref{fig:stageccds} reveals a nontrivial fraction of models in Stages II and III that exhibit very red colors. In Stages 0 and I, redness resulting from the combination of extinction and excess emission from the circumstellar envelope is expected, but this envelope is essentially absent from later stages by definition. Red colors in Stage III are instead caused by additional emission from heated dust in the ambient medium present in some models (see Section \ref{sec:2.1.4} for properties), in turn due to high energy input from extremely luminous sources. Reddening for Stage II models is mostly caused by emission and extinction by the disk, but models with hot sources can also be reddened by a heated medium along sightlines that do not go through the disk. These make up the tail feature in the Stage II panel of Figure \ref{fig:stageccds}. Most of the red Stage III models come from the \texttt{s---smi} geometry, but not all; edge cases are discussed later in this section.

In the models, dust is destroyed by sublimation at 1600 K. However, this is the only modeled pathway for dust destruction; no other mechanisms are implemented by R17 or this work. The floor dust density of $10^{-23}$ g cm$^{-3}$ is likely denser than what might be expected in the vicinity of a bare high-temperature star. While some Stage III YSOs could theoretically exhibit some redness via this phenomenon of heated dust, observing such an object is unlikely. These models, which consist of the combined SEDs of a hot star and heated dust, could see an alternate use as model contaminants; for example, main-sequence stars traveling through a dusty medium. Some additional discussion of reddening by medium emission is contained in Appendix \ref{ap:ccds}.

Our scheme for determining stages has some quirks when interacting with the models. The definition of Stage III includes models with no non-ambient density structures, which is nominally only two of the eighteen available geometries (\texttt{s---s-i} and \texttt{s---smi}). However, geometries with an ambient medium--including all geometries with envelopes and half of the others (bare star and star + disk only)--can be Stage III under particular circumstances. The medium is intended as a lower limit to density in models that have it. Some Stage III models therefore ostensibly have envelopes and/or disks with assigned properties, but they are treated as nonexistent by the radiative transfer because they fail to rise above the floor density of the ambient medium at any point, so they are functionally bare PMS stars with a medium. Similarly, the randomly sampled nature of the models can cause models with disks (with or without a medium) to have disks with larger inner than outer radii. In such a case, the disk is not created, and the model is run as if there were no disk. Should the disk be the only non-ambient density structure, these models therefore become part of Stage III.

Stagification also produces models with no defined stage ($\approx$ 11\% of all models). Models with a class but no stage have envelopes with less than 0.1 $M_\odot$ of mass, but no disk. These can be useful in portraying emission from cores at long wavelengths, but are not assigned a stage because they do not conform to our definition and are generally not accounted for in any theory of star formation. Models with neither a class nor a stage are incomplete (again, see Section \ref{sec:3.6}).

E09 contains an alternate definition for YSO evolutionary stages, put forth in R06. In this definition, boundaries between stages are set by the ratio of envelope infall rate to stellar mass ($\dot{M}_{\rm env}/M_\star$) and the ratio of disk mass to stellar mass ($M_{\rm disk}/M_\star$). We do not make use of this definition; applying it requires knowledge about the mass and accretion rate of the central source of each YSO, which R17 purposefully does not include.

\subsubsection{Comparison}\label{sec:4.2.3}
Using the previous sections, we compare the distributions of models in color space by class and by stage. While there is overlap between the two, the evolution in class generally tends monotonically blueward to zero color while the evolution in stage tends to push redward before returning to zero. While models of a particular class often coexist in color space with their corresponding stage, these coexistent areas are also often occupied by other classes and stages. Semi-reliable distinction can be achieved by including longer-wavelength (e.g. F2550W) emission driven by the presence of warm disks. The degree to which YSOs of different classes and stages may be identified and distinguished using their position in IR color space is therefore qualitatively in line with R06 and G09.

\begin{figure}
    \centering
    \includegraphics[width=0.99\textwidth]{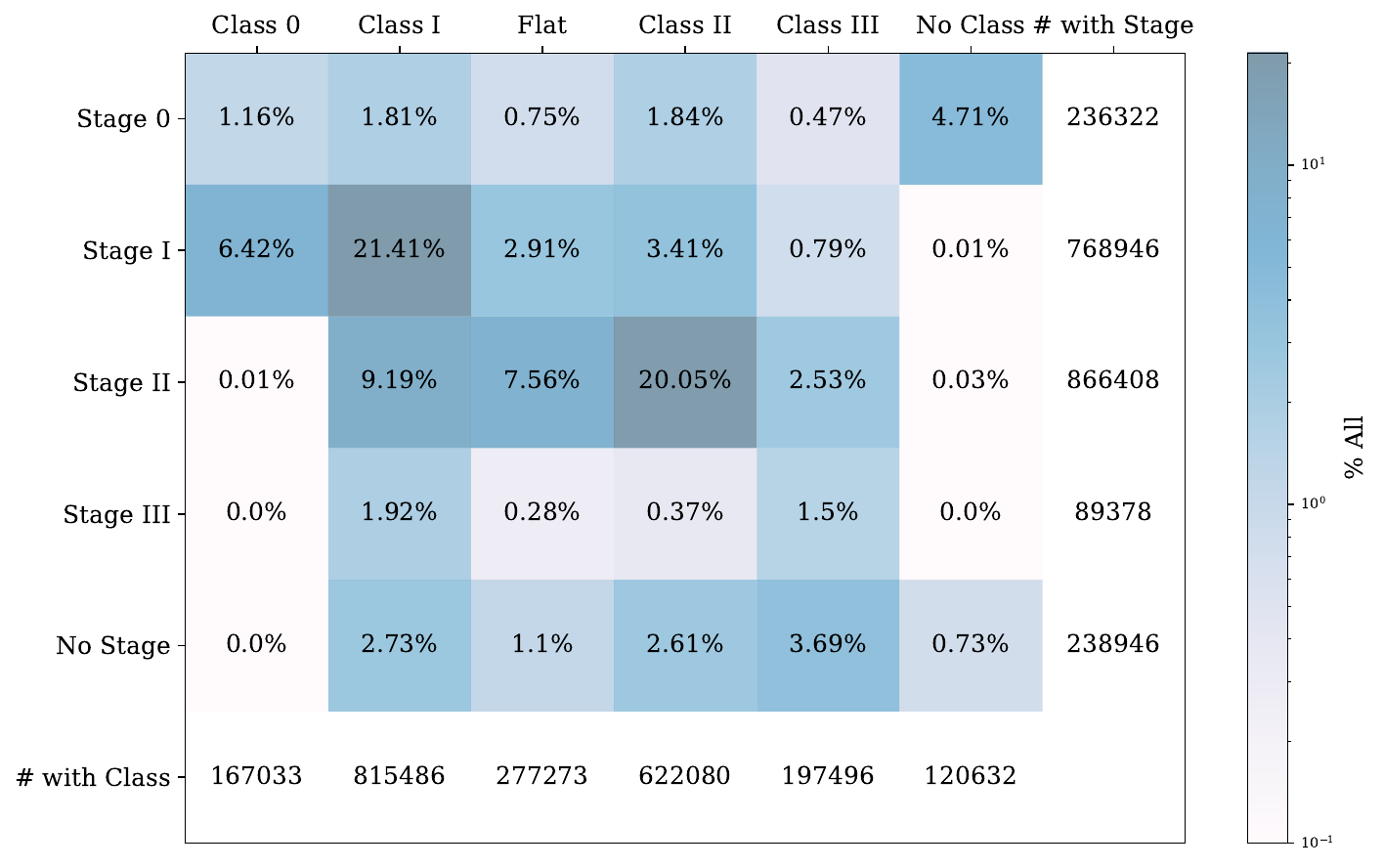}
    \caption{Confusion matrix for class and stage. Each cell displays the amount of models that have a given class and a given stage as a percentage of the total number of available models. Classes are assigned using the SED in the largest aperture available to the models, 10$^6$ AU. The total number of models that have a given stage (across all classes) and class (across all stages) are also displayed. ``No class" models are a mixture of optically thick models and incomplete SEDs. ``No stage" models exist outside of our definition (see \S\ref{sec:4.2.2}) or are incomplete.}
    \label{fig:confusion}
\end{figure}

In addition to comparing class and stage observationally, we can evaluate the extent to which the concepts themselves are related. In Figure \ref{fig:confusion}, we show a confusion matrix for class and stage. There are clear correlations between class and stage. Majorities of Class I and Class II models are also Stage I and Stage II, respectively, and there is a clear shift towards later stages as class increases in general. However, models of a given spectral class are capable of being many different stages, and vice versa. Approximately a third of Class I sources have a stage other than I, more than a quarter of Class II sources have a stage other than II, and the majority of Class III sources are not Stage III.

This confusion illustrates the magnitude of the effects of confounding factors. Stage II models, for example, have only disks by definition. However, they would be misidentified as having envelopes (i.e. classified as I/Flat) about as often as they would be classified ``correctly" if these numbers are taken at face value. They could also be misidentified as bare stars a non-negligible amount of the time. The disjunction between class and stage is driven by the viewing angle, which causes edge-on models to appear to have earlier spectral classes and face-on disks/models to appear to have later ones, and by foreground extinction from the models' ambient medium. Effects from the latter can also be seen in Stage III models, which have no envelope or disk by definition, and yet are likely to have a class that is not III due to the presence of an ambient medium.

We have used our models to outline the distinction between observational class and evolutionary stage. However, we caution against over-interpretation of these results. The number of models in each geometry of R17 is determined by model complexity rather than its ability to accurately represent evolutionary stages or observational classes. The distribution of models within the classes and stages in Figure \ref{fig:confusion} (the numbers in ``\# with Class" and ``\# with Stage") is consequently not representative of the relative fraction of these sources in real star-forming clouds. We also reiterate that all models have randomly sampled properties and intentionally do not conform to any single model of protostellar evolution. The results from this section therefore include SEDs from models that may not occur in nature, and the degree to which the confusion matrix may be impacted by model feasibility has not been quantified. (We expect the conclusions drawn from color-color diagrams to be minimally affected by model feasibility given their qualitative agreement with the literature; see Appendix \ref{ap:ccds} for a direct comparison to R06.) We also acknowledge that our method for assigning evolutionary stages to our YSOs is distinct from schemes used by other works. Our definition is based on commonly used distinctions between stages where possible and relies on accepted physics for any further additions. We therefore expect it to sort models appropriately, but assigned stages may differ depending on the scheme.

\section{Conclusion}\label{sec:5}
In this paper, we have presented a significant update to the \citet{robitaille2017} set of YSO models. We have calculated several quantities using the existing model parameters and infrastructure. Through these calculations, we have substantially expanded the utility of the existing models within contexts where they are currently in use. Models from the original set are commonly used as template SEDs to determine YSO properties. Our additional content increases the number of quantities that may be constrained through this method, including:
\begin{itemize}
    \item the ``observed" mass of a core within an aperture (``Line-of-Sight Masses'')
    \item the ``actual" mass present around the source (``Sphere Masses'')
    \item the average dust temperature (``Sphere Mass-Weighted Temperatures, Line-of-Sight Mass/Photon-Weighted Temperatures'')
    \item the extinction as a result of circumstellar dust (``Av'')
\end{itemize}
all of which provide a more complete characterization of an observed YSO. We have also provided further insight into the physical state of each model by calculating a baseline for the stability of the disk for models with disks as well as the average dust temperatures within the series of apertures in which the SEDs are defined.

Beyond our work's application to existing use cases, we have utilized our new additions to derive new results previously inaccessible to users of these models. In particular, we use our updates to provide guidance for measuring the mass of an optically thin core based on the observed flux. We find that assuming a constant dust temperature may yield an inferred mass within a factor of two of the correct value. However, that level of precision requires assuming a representative dust temperature. The ``correct" temperature is often not the canonical 20 K, and varies with the observed flux and aperture size. Moreover, the inferred mass may be underestimated by over a factor of two if the dust being observed is insufficiently optically thin.

Our newly calculated properties also enable us to assign each model an observational class and evolutionary stage. We use our convolved SEDs (now including every filter on the James Webb Space Telescope) to locate each model in IR color space and map out the regions that are home to each class and stage. We find that mid-IR colors are sensitive to the presence of disks, making them a useful tool to identify YSOs and distinguish between evolutionary stages. Further, the expanded scope and larger size of this set enables users to probe other color spaces and set expectations for newly obtained data. We also use these assigned classes and stages to evaluate the extent to which the concepts of ``class" and ``stage" are related in our set of models. We find a correlation between the two, as expected, but also find that the mapping from class to stage can be confused by viewing angle and foreground extinction in a nontrivial fraction of cases.

The models and associated scripts have been made publicly available at \url{https://doi.org/10.5281/zenodo.8114592}.
%and you best prepare for some HEAT on the next one
\acknowledgments
AG acknowledges support from the NSF under grants AST 2008101 and CAREER 2142300. This research has made use of the Spanish Virtual Observatory (\url{https://svo.cab.inta-csic.es}) project funded by MCIN/AEI/10.13039/501100011033/ through grant PID2020-112949GB-I00. We thank the anonymous referee for providing comments that were instrumental in improving the paper and data products.

\software{Astropy \citep{astropy:2013, astropy:2018}, Hyperion \citep{robitaille2011}, Matplotlib \citep{matplotlib}, NumPy \citep{numpy}, SciPy \citep{scipy}.}

\bibliography{ref.bib}
\bibliographystyle{aasjournal.bst}

\appendix
\section{Luminosities}\label{ap:lum}
\begin{figure}
    \centering
    \includegraphics[width=0.99\textwidth]{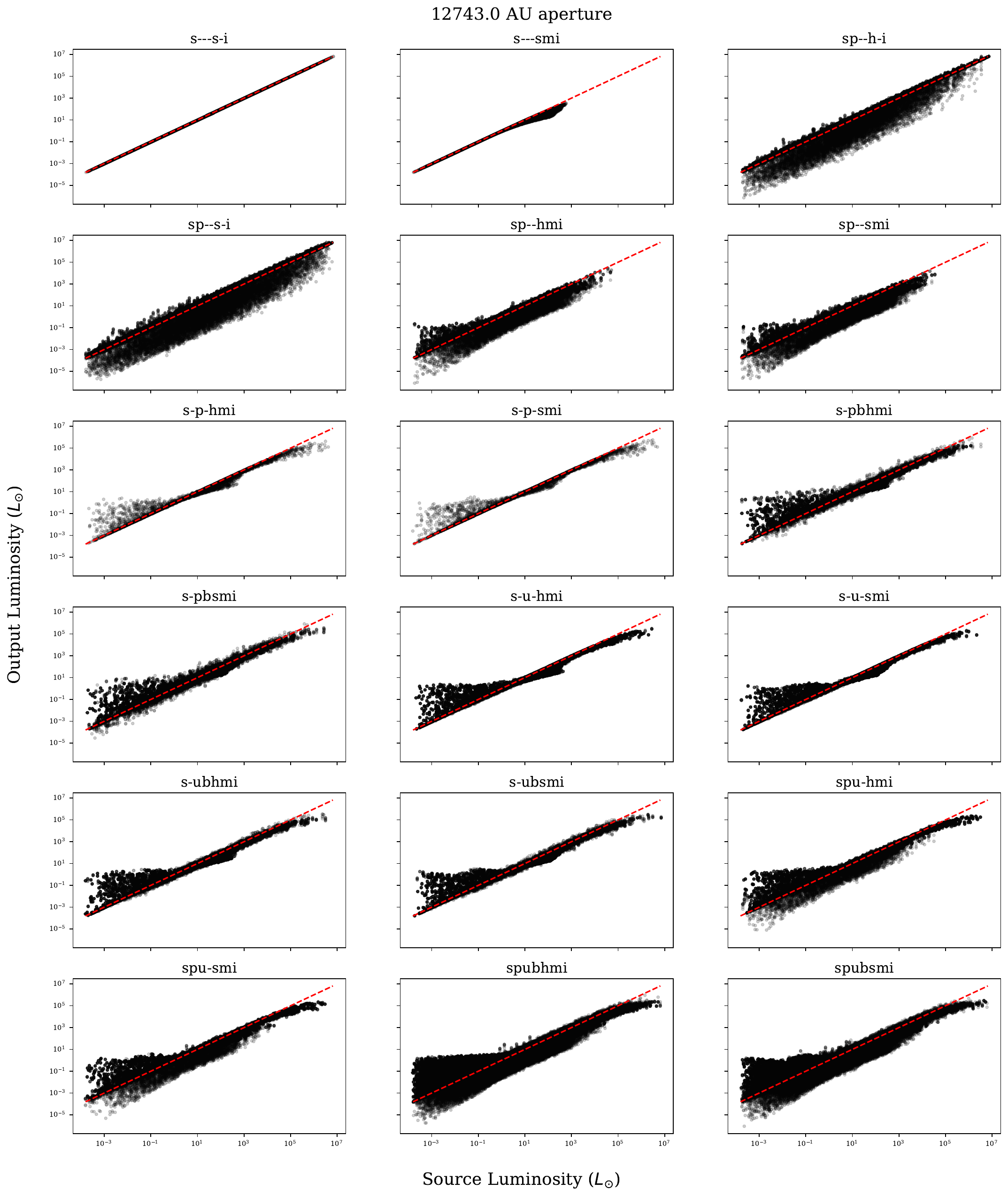}
    \caption{Luminosities recovered from every model SED using Eq. \eqref{eq:projection} plotted against the luminosities of the sources at the center of each model. A 1-1 line is plotted in red. All SEDs are observed in an aperture of radius $\approx$10,000 AU at a distance of 1 kpc. For SEDs with multiple viewing angles, recovery is done for each independently.}
    \label{fig:lums}
\end{figure}

\begin{figure}
    \centering
    \includegraphics[width=0.97\textwidth]{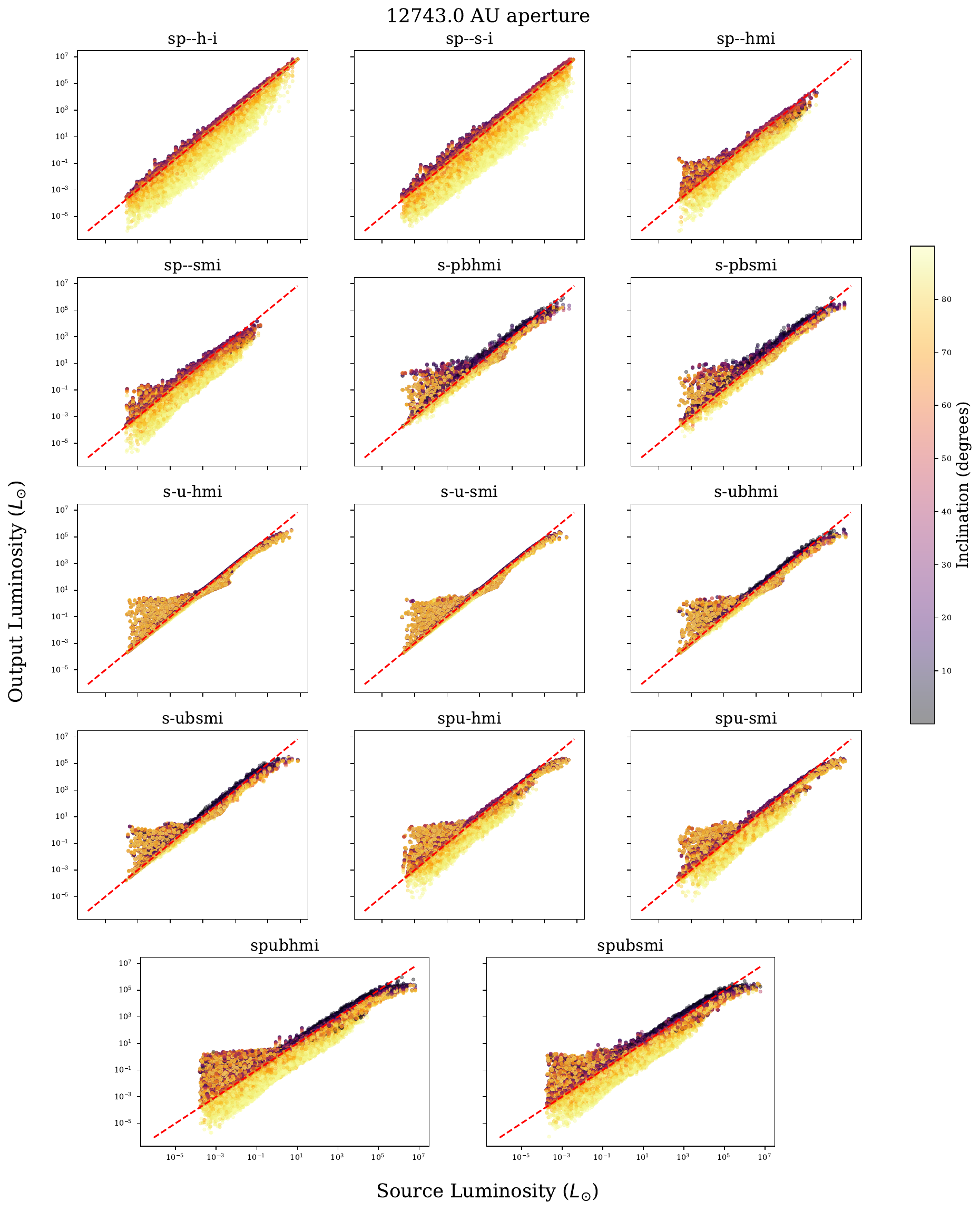}
    \caption{The same as Figure \ref{fig:lums}, but limited to models with $\theta$ dependence and colored by the viewing angle (see \S\ref{sec:2.3} for more details.)}
    \label{fig:lums_inc}
\end{figure}

\begin{figure}
    \centering
    \includegraphics[width=0.99\textwidth]{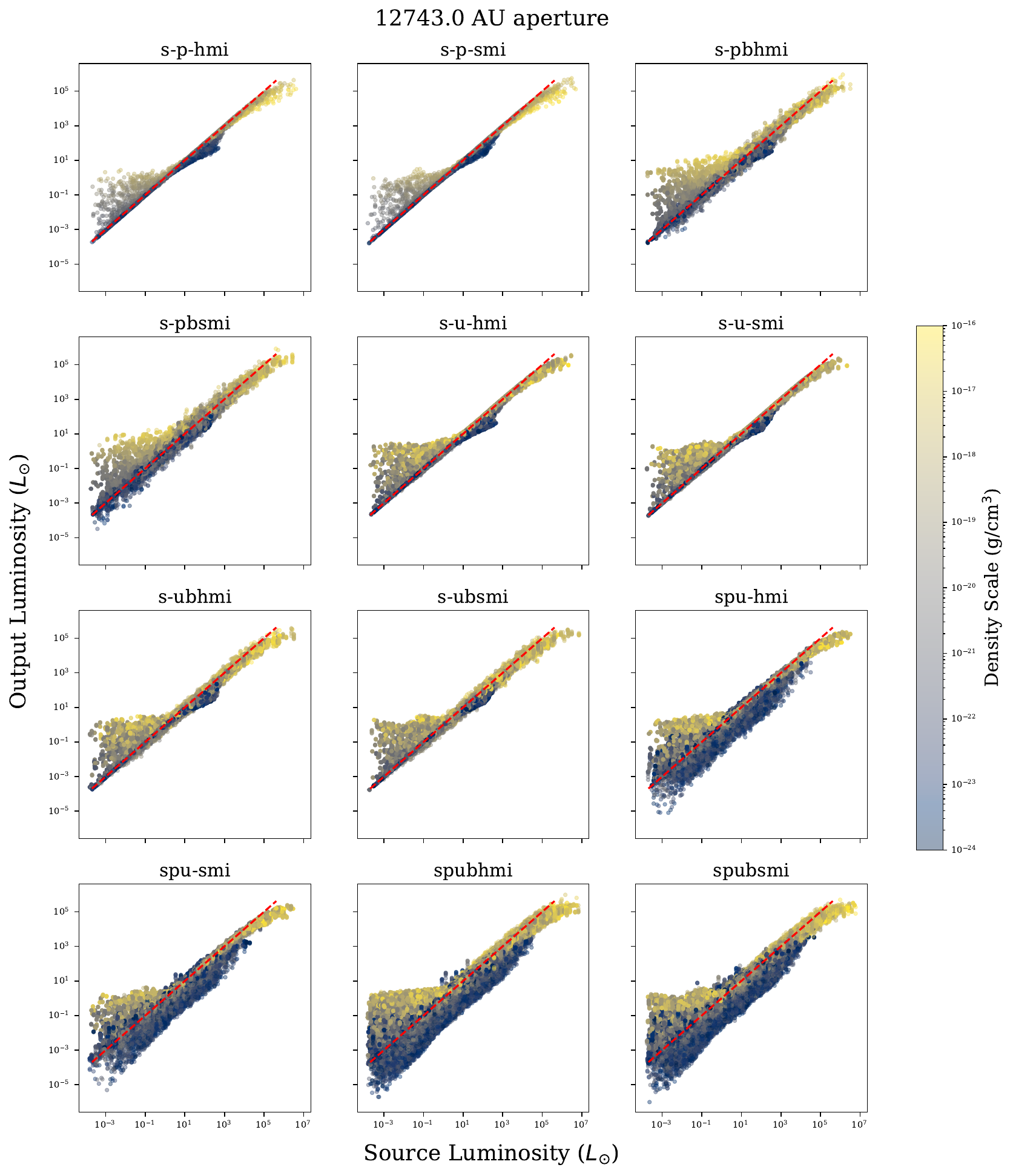}
    \caption{The same as Figure \ref{fig:lums}, but limited to models with envelopes and colored by the density scale of the envelope (see \S\ref{sec:2.1.1} for more details).}
    \label{fig:lums_env}
\end{figure}

We compare the luminosity of the central source in each YSO model, calculated directly from the model parameters, to the isotropic luminosity projected from the flux density calculated by Hyperion. In effect, this allows us to evaluate whether energy emitted by the source is conserved. Luminosity is recovered with the equation:
\begin{equation}
    L=4\pi d^2 \int S_{\nu}d\nu
    \label{eq:projection}
\end{equation}
where $d$ is the distance to the source, here 1 kpc. Results for a 10,000 AU aperture are collected in Figure \ref{fig:lums}; we color these plots by viewing angle for models that depend on it in Figure \ref{fig:lums_inc} and density scale for models with envelopes in Figure \ref{fig:lums_env} to highlight the dependence of observed features on these properties. For the geometry with no density structures, we successfully recover the source luminosity. Geometries with dust deviate from the source due to contributions from dust and observational effects.

Many geometries experience a spread in recovered luminosity regardless of source brightness. This spread occurs in all geometries that have a $\theta$ dependence--rotational flattening, cavities, etc.--and is particularly true for disk-only models, where the largest spreads can be seen (\textit{ex. Figure \ref{fig:lums}, top right}). This is a consequence of viewing angle; light traveling from the source is scattered out of the denser edge-on lines of sight and into the less dense face-on angles, resulting in a respective deficit/excess in flux. This inclination dependence is highlighted in Figure \ref{fig:lums_inc}.

Many model geometries also exhibit over-luminous ``wedges" from $\approx$10$^{-4}$ - 10$^{-1}$ $L_\odot$; this can be seen clearly in Figure \ref{fig:lums} (\textit{bottom row, middle panel}). Figure \ref{fig:lums_env} shows that models with higher dust density scales generally have higher recovered luminosities, indicating that the heightened luminosities result from emission by heated dust surrounding the central source--the more dust surrounding the source, the greater the discrepancy can be.

Finally, the introduction of an ambient medium results in the appearance of under-luminous ``dip" features driven by extinction. Such a dip can be seen at $\approx$1 $L_\odot$, visualized clearly in Figure \ref{fig:lums} (\textit{top row, middle panel}). These features are persistent across geometries and increasingly pronounced with aperture size as more dust, whether from an envelope or the ambient medium, is included in the aperture. High-luminosity models experience a second dip beginning at $\approx$ 10$^4$ $L_\odot$ as models with more heated dust (due to higher input luminosity and density scale) also experience extinction from the ambient medium, which is showcased in Figure \ref{fig:lums_env}. This higher-luminosity dip is a feature that is present by virtue of having widespread dust, but is reduced or absent in some panels of Figure \ref{fig:lums} due to post-processing for S/N; see Section 4.2.4 of R17 for details.

\section{Additional color-color diagrams}\label{ap:ccds}
\begin{figure}
    \centering
    \includegraphics[width=0.99\textwidth]{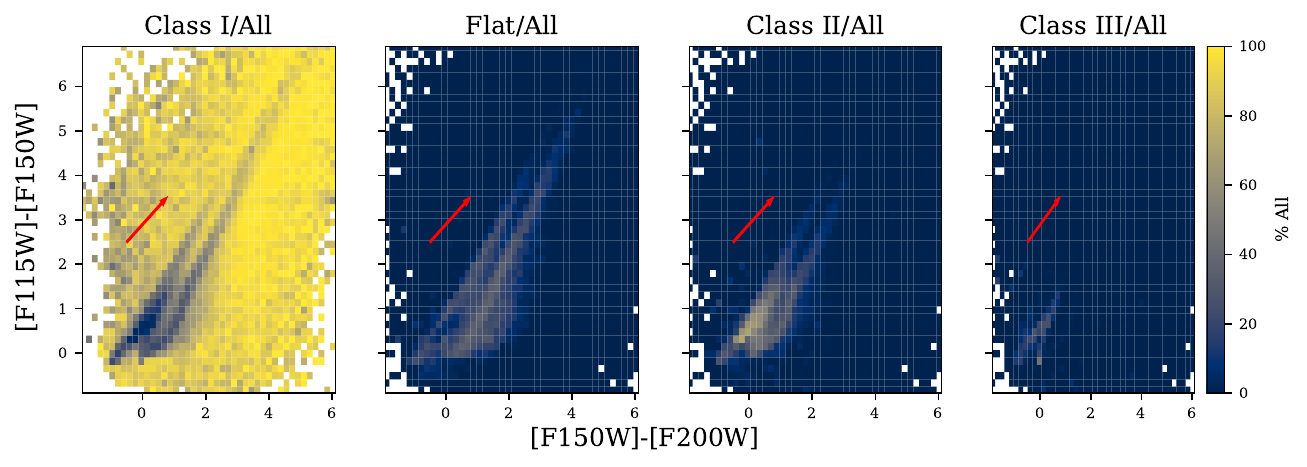}
    \includegraphics[width=0.99\textwidth]{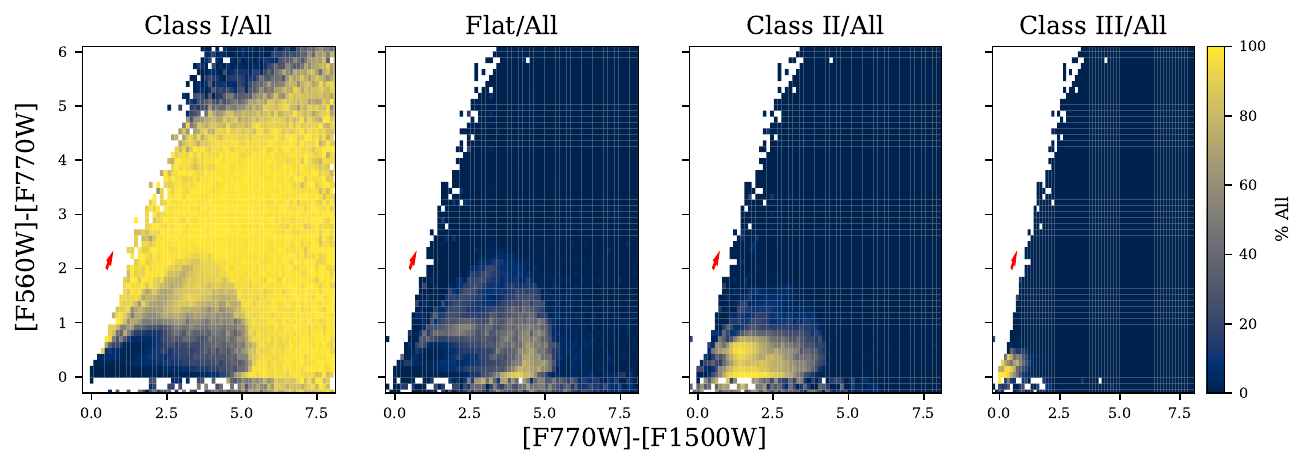}
    \includegraphics[width=0.99\textwidth]{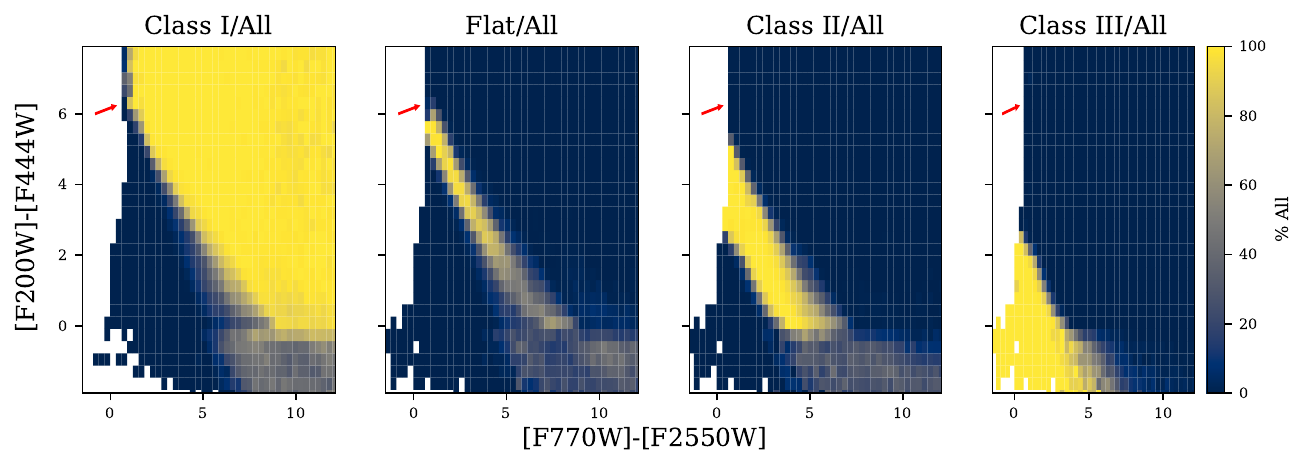}
    \caption{2D histograms of the models in the same color-color spaces as in Figure \ref{fig:classccds}, but colored by the number of models that occupy each bin as a percentage of the total. (The more yellow a bin is, the larger the share of all models in that bin with that class.) See \S\ref{sec:4.2.1} for details on classification.}
    \label{fig:class_frac}
\end{figure}

\begin{figure}
    \centering
    \includegraphics[width=0.99\textwidth]{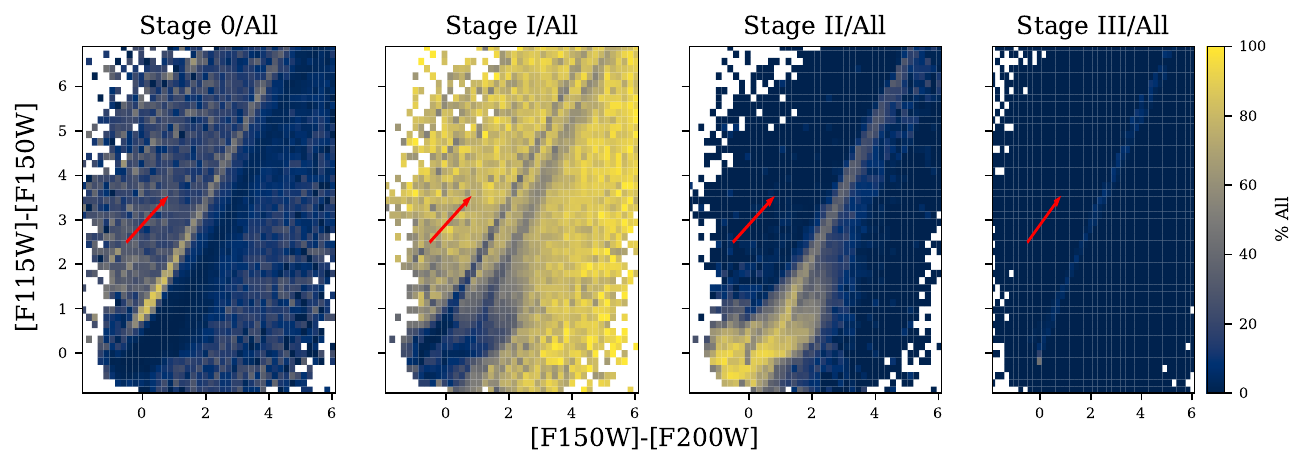}
    \includegraphics[width=0.99\textwidth]{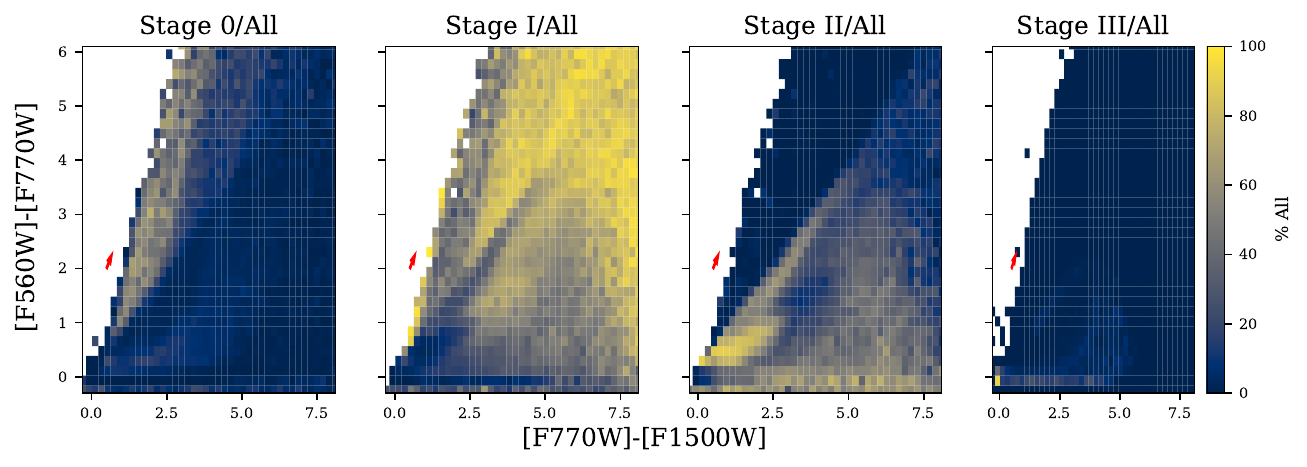}
    \includegraphics[width=0.99\textwidth]{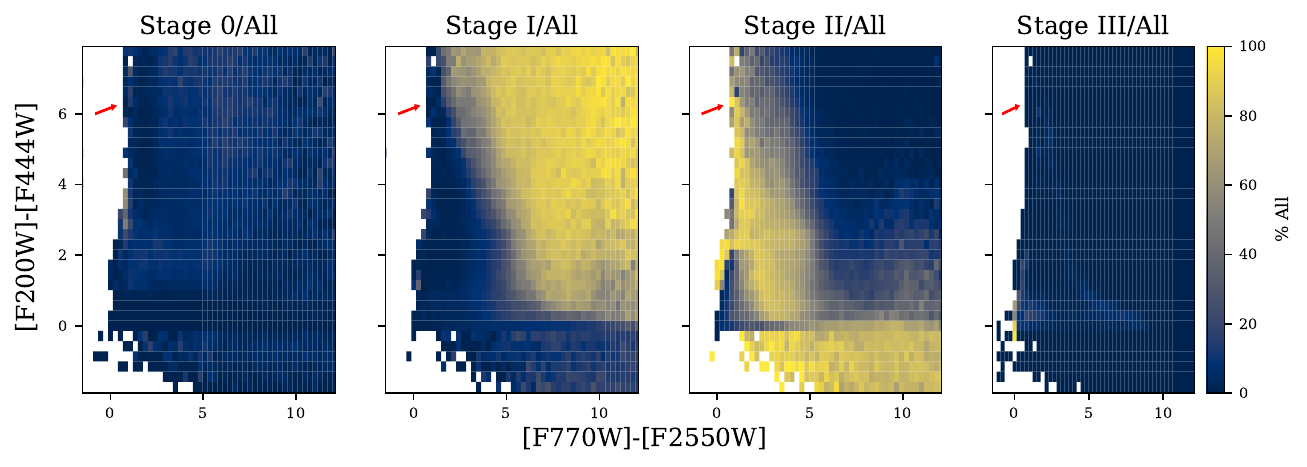}
    \caption{The same as Figure \ref{fig:class_frac}, but separated by evolutionary stage instead. See \S\ref{sec:4.2.2} for details on stage assignment.}
    \label{fig:stage_frac}
\end{figure}

In Sections \ref{sec:4.2.1} and \ref{sec:4.2.2}, we separated the models by class and stage and visualized them in JWST color space. In this appendix, we return to Figures \ref{fig:classccds} and \ref{fig:stageccds} but instead plot the number of models of each class and stage as a percentage of the total in Figures \ref{fig:class_frac} and \ref{fig:stage_frac}. These plots generally reinforce observations made using the non-percentage versions. Class I objects are very red, and successive classes become less so. Successive stages, however, trace out a more complicated (and distinct) path in color space, with disk-only models tending to be preferentially reddened in colors involving mid-IR wavelengths compared to models that are either envelope-dominated or do not have disks.

\begin{figure}
    \centering
    \includegraphics[width=0.49\textwidth]{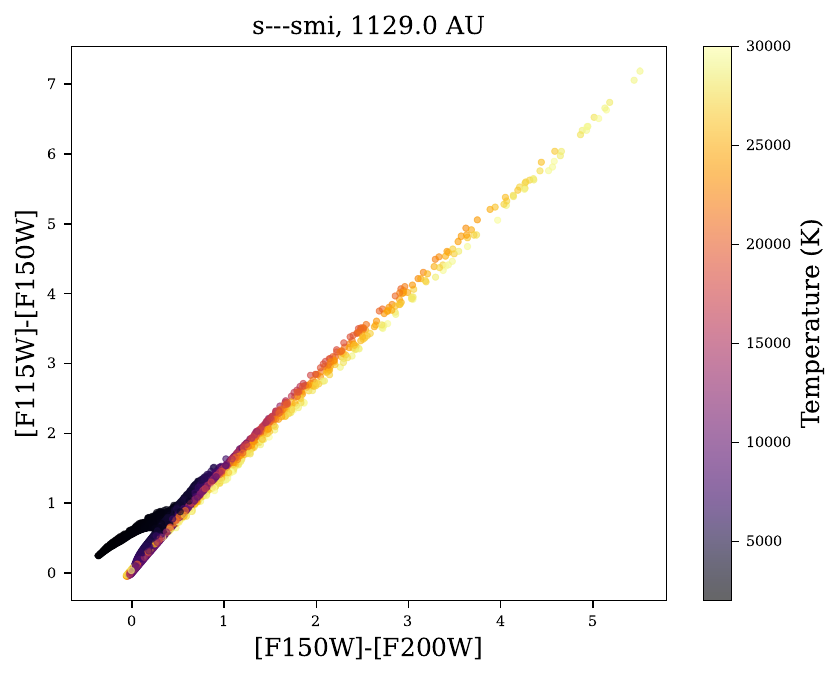}
    \includegraphics[width=0.49\textwidth]{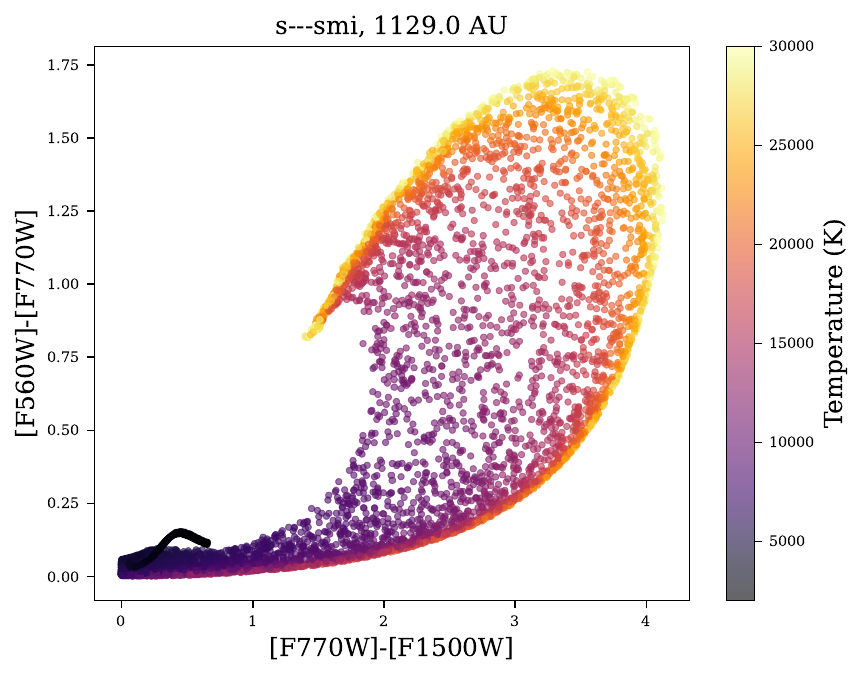}
    \caption{NIRCam and MIRI color-color diagrams for the \texttt{s---smi} geometry (akin to Stage III from Figure \ref{fig:stageccds}) colored by the temperature of the source in each model. Models with hotter sources are redder than those with colder sources in both color spaces.}
    \label{fig:mediumonly}
\end{figure}

\begin{figure}
    \centering
    \includegraphics[width=0.49\textwidth]{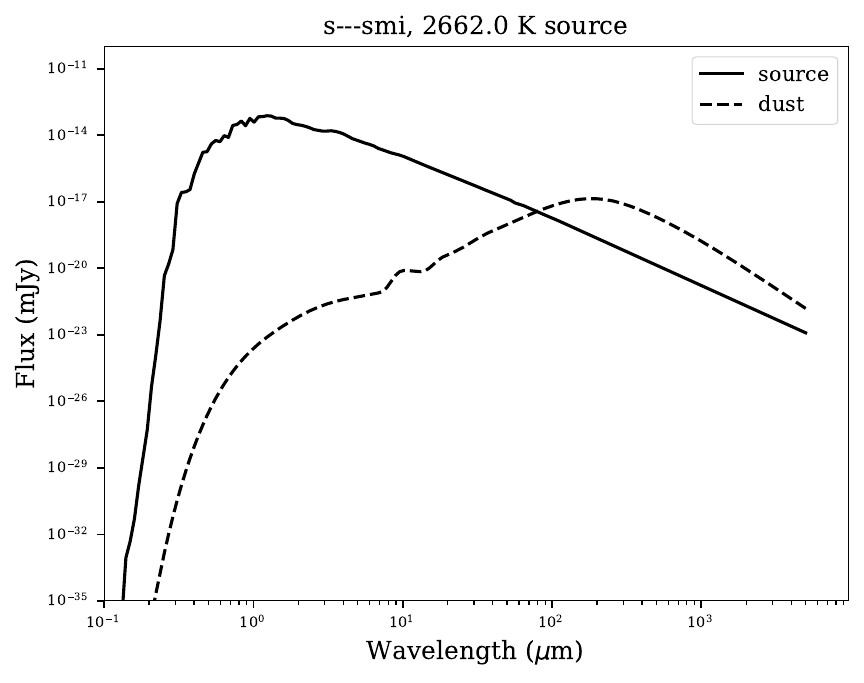}
    \includegraphics[width=0.49\textwidth]{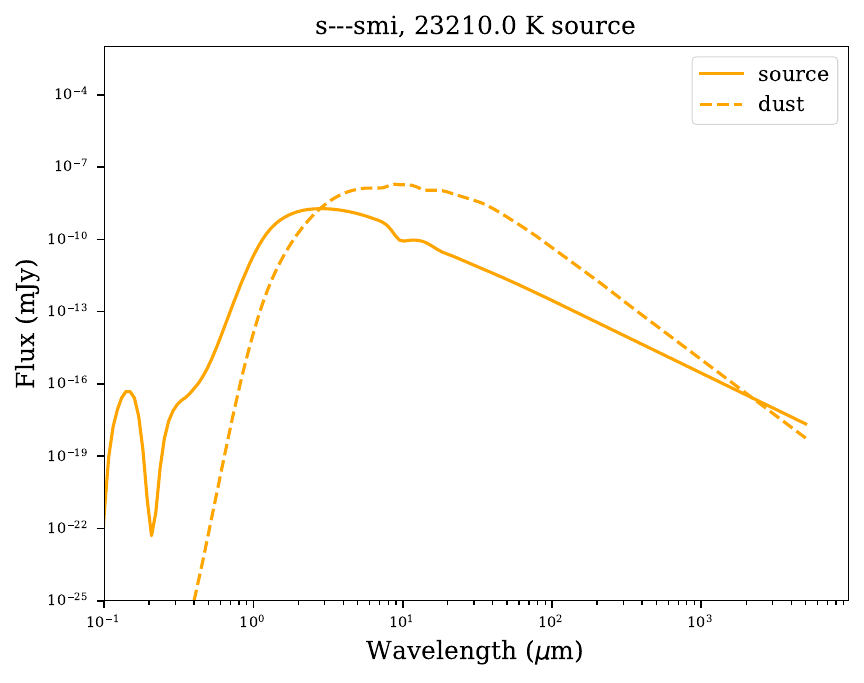}
    \caption{SEDs from the \texttt{s---smi} geometry for a model with a cold source (\textit{left}) and a hot source (\textit{right}), broken down by component. The dust around the hot source is clearly dominant in the near- and mid-IR, unlike that around the cold source. The viewing aperture for both SEDs is chosen to be as close to the aperture used in Figure \ref{fig:mediumonly} as possible.}
    \label{fig:sourcecomp}
\end{figure}

In Figure \ref{fig:stageccds}, noteworthy fractions of the models in Stages II and III exhibit red colors. We attribute this redness to additional emission from the ambient medium driven by exposed sources with high luminosity. To isolate this effect, we consider the \texttt{s---smi} geometry, which is composed solely of models with a source and an ambient medium. In Figure \ref{fig:mediumonly}, we illustrate the dependence of position in NIRCam and MIRI color space on source temperature. There is a clear correlation between redness and temperature, and this medium-only geometry reproduces the shape of the Stage III models in Figure \ref{fig:stageccds}.
To determine the cause of this redness, we compare the SEDs of \texttt{s---smi} models with cold sources to models with hot sources. In general, we find that the medium around hot sources dominates the NIR/MIR range of the SED, while colder sources tend to be brighter than their surrounding medium over the same range. Examples are plotted in Figure \ref{fig:sourcecomp}.
Dust emission from the medium around hot sources therefore results in more IR flux, which in turn causes these models to appear redder.

\begin{figure}
    \centering
    \includegraphics[width=0.99\textwidth]{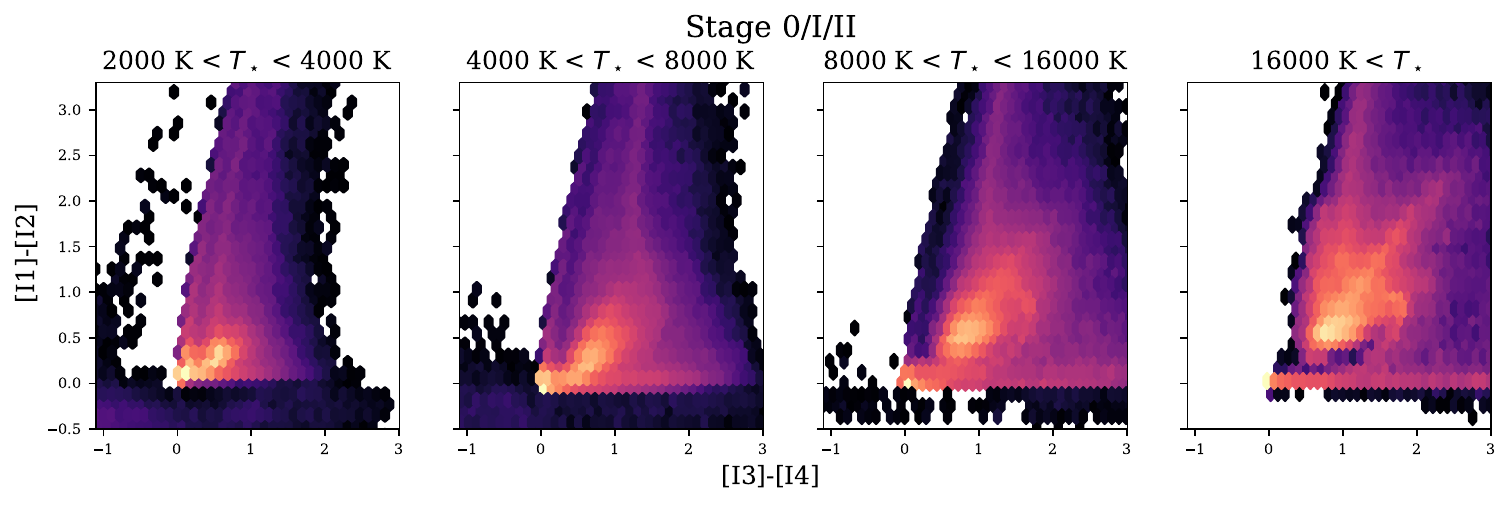}
    \includegraphics[width=0.99\textwidth]{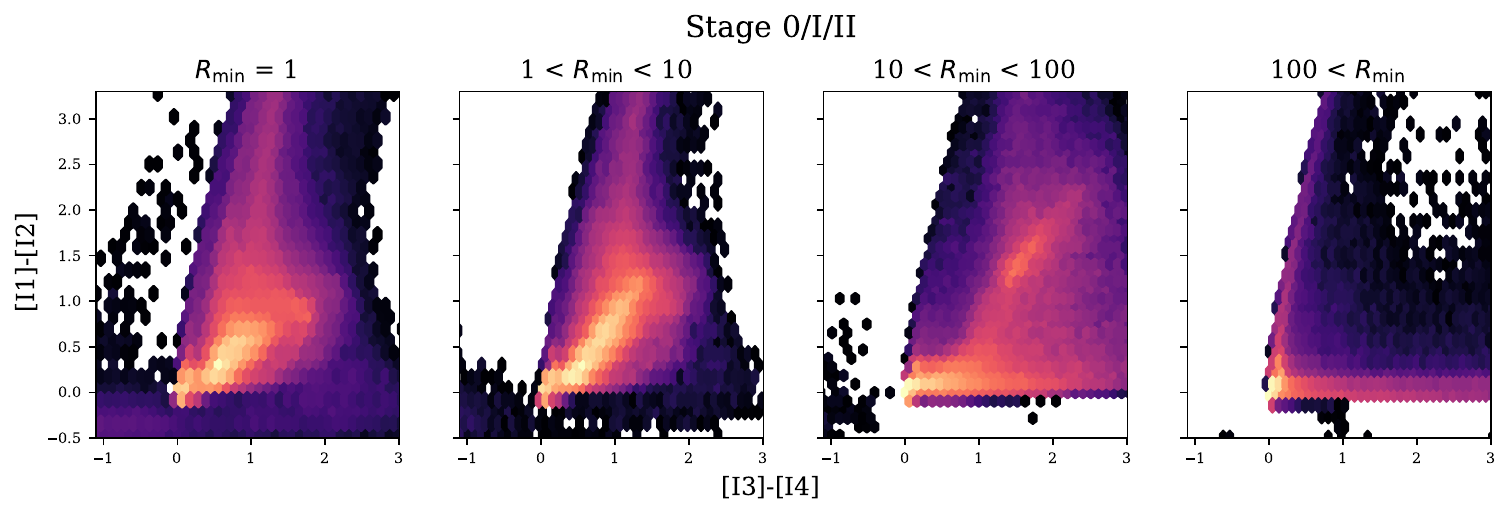}
    \caption{Plots of our models in IRAC color-color space, separated by stellar temperature (\textit{top}) and inner radius of circumstellar envelope/disk, in terms of dust sublimation radius (\textit{bottom}). These are analogous to  Figures 26-28 in R06, which separates its models in a similar way. All models that fall into Stages 0, I, and II (see \S\ref{sec:4.2.2}) are included in both plots.}
    \label{fig:r06_redo}
\end{figure}

Between Section \ref{sec:4.2} and this appendix, we have mapped out the positions of our models in JWST color-color space, as well as how those positions change as a function of spectral class and evolutionary stage. To demonstrate that our results are consistent with previous work, we reproduce plots from R06 using our updated model set. In addition to evolutionary stage, R06 breaks its grid down in color space by properties intrinsic to each model: the envelope accretion rate, inner radius of a model's envelope/disk (in terms of dust sublimation radius), and the temperature of the central source. Our set of models can also be separated by source temperature and inner radius in the same way as in R06, as these parameters are common to both sets of models. We recreate these plots in Figure \ref{fig:r06_redo}.

We find general agreement between our results and those of R06. Models with higher temperatures are redder on average, in accordance with R06's generalization of the findings of \citet{whitney2004} to near- and mid-IR wavelengths. As the inner radius of our models increases, there is a clear evolution from being preferentially reddened in [I1]-[I2] to [I3]-[I4], to the point where most models in the largest radius bin have a color close to zero in [I1]-[I2]. A similar phenomenon occurs in R06; it is explained there as a result of a shift in flux from shorter to longer wavelengths as the dust temperature around the source decreases. As this shift occurs, the shorter-wavelength fluxes increasingly become solely due to stellar photospheres, which means that the models exhibit photospheric colors (which tend close to zero). In general, our results have more spread in color-color space than their counterparts in R06, likely due to the increased size and randomly sampled nature of the models. However, it is clear that despite this spread, the behavior of our model set comports with that of the R06 grid.
\end{document}